\documentclass[11pt]{article}

\usepackage[title,toc,titletoc]{appendix}
\DeclareMathAlphabet{\mathlcal}{U}{dutchcal}{m}{n}
\SetMathAlphabet{\mathlcal}{bold}{U}{dutchcal}{b}{n}
\usepackage{IEEEtrantools}
\usepackage{etoolbox}
\patchcmd{\appendices}{\quad}{. }{}{}
\usepackage{graphicx} 
\usepackage{subfiles}
\usepackage{float}
\usepackage[a4paper, total={6in, 8in}]{geometry}
\usepackage{color,soul}
\usepackage{verbatim}
\usepackage{amsthm}
\usepackage{amsmath,amssymb,tabu}
\usepackage{amsfonts}
\usepackage{enumitem}
\usepackage[english]{babel}
\usepackage{mathrsfs}  
\usepackage{mathtools}
\usepackage[mathscr]{euscript}
\usepackage{comment}
\usepackage{aliascnt}
\usepackage{bbm}
\usepackage{tikz}
\usepackage{subfig}
\def\bs{\ensuremath\boldsymbol}

\def\vp{\ensuremath\varepsilon}

\allowdisplaybreaks

\newtheorem{proposition}{\textbf{{Proposition}}}
\newtheorem*{proposition*}{\textbf{{Proposition ID}}}
\newtheorem{theorem}{Theorem}

\makeatletter
\@addtoreset{theorem}{section}
\makeatother

\newtheorem{Assumption}{\textbf{{Assumption}}}
\newtheorem*{Assumption*}{\textbf{{Assumption ID}}}
\usepackage{hyperref}
\usepackage{natbib}
\usepackage[nameinlink]{cleveref}
\DeclareMathOperator*{\argmax}{arg\,max}

\DeclareMathOperator*{\argmin}{arg\,min}
\usepackage{float}

\newenvironment{pff}[1][Proof]{\vspace{1ex}{\noindent\textbf{#1.} }\hspace{.1em}}
{\hfill\qed\vspace{1ex}}
\makeatletter
\newcommand{\pushright}[1]{\ifmeasuring@#1\else\omit\hfill$\displaystyle#1$\fi\ignorespaces}
\newcommand{\pushleft}[1]{\ifmeasuring@#1\else\omit$\displaystyle#1$\hfill\fi\ignorespaces}
\makeatother
\newcommand{\bcdot}{\raisebox{1pt}{\textbf{\large .}}}
\makeatletter
\def\blfootnote{\xdef\@thefnmark{}\@footnotetext}
\makeatother
\usepackage{adjustbox}

\usepackage{multirow}
\usepackage{xr}
\begin{document}
	
%%%%%%%%%%    Title    %%%%%%%%%%
\title{\textbf{Shrinkage Estimation of Network Spillovers with Factor Structured Errors}$^*$}
\date{\vspace{-5ex}}
\maketitle
\vspace{-40pt}
\begin{center}
\begin{tabular}
	[c]{ccc}%
	Ayden Higgins$^{\dagger}$  & \hspace{0.54cm} & Federico Martellosio$^{\ddagger}$\\
	University of Surrey, UK &  & University of Surrey, UK \\
\end{tabular}\\
\vspace{10pt}
\date{\today}
\vspace{10pt}
\end{center}
	
%%%%%%%%%%    Abstract    %%%%%%%%%%	
\begin{abstract}
This paper explores the estimation of a panel data model with cross-sectional interaction that is flexible both in its approach to specifying the network of connections between cross-sectional units, and in controlling for unobserved heterogeneity. It is assumed that there are different sources of information available on a network, which can be represented in the form of multiple weights matrices. These matrices may reflect observed links, different measures of connectivity, groupings or other network structures, and the number of matrices may be increasing with sample size. A penalised quasi-maximum likelihood estimator is proposed which aims to alleviate the risk of network misspecification by shrinking the coefficients of irrelevant weights matrices to exactly zero. Moreover, controlling for unobserved factors in estimation provides a safeguard against the misspecification that might arise from unobserved heterogeneity. The asymptotic properties of the estimator are derived in a framework where the true value of each parameter remains fixed as the total number of parameters increases. A Monte Carlo simulation is used to assess finite sample performance, and in an empirical application the method is applied to study the prevalence of network spillovers in determining growth rates across countries.
\color{white}.\color{black}
\end{abstract}		
\vspace{-0.1in}

\blfootnote{\hspace{-0.2in}$^*$We are grateful to three anonymous referees and to the Associate Editor for insightful comments that helped us to improve the paper. We would also like to thank Xun Lu and Liangjun Su for sharing their data with us, as well as  
Valentina Corradi, Jo$\tilde{\text{a}}$o Santos Silva, Sorawoot Srisuma and Martin Weidner for their comments and suggestions.}
\blfootnote{\hspace{-0.2in}$^{\dagger}$Email: a.higgins@surrey.ac.uk}
\blfootnote{\hspace{-0.2in}$^{\ddagger}$Email: f.martellosio@surrey.ac.uk}

%%%%%%%%%%    Main Text    %%%%%%%%%%%
\newpage 

\section{Introduction}

\label{intro}
Increased attention is being given over to panel data models which take into account cross-sectional interaction. These models have proven to be empirically relevant in a diverse range of economic settings, such as social interactions between individuals, business connections between firms, trading relations between nations, and dependencies between financial assets. At the heart of many econometric models of this kind lies a weights matrix, which summarises the network of connections between interacting cross-sectional units. Yet networks are rarely fully observed, and the uncertainty in how a weights matrix should be specified has been a common critique of this growing literature \citep[see, e.g.,][]{blume,de_paula_recovering_nodate,lewbel_2019}. In practice, situations in which networks are partially observed are more frequent, with some information being available on cross-sectional links, or their absence, as well as information on other network structures such as groupings. As an example, within a school one might observe family, friendship, classroom and cohort groupings, each of which provide information on the network of connections between different students. 
In other settings, such as international networks, there are multiple ways to quantify connectivity between nations, including economic measures such as trade and foreign direct investment flows, physical distance, and infrastructure links. Nevertheless, it is not usually obvious how these pieces of the jigsaw fit together, and this uncertainty inevitably increases the risk of model misspecification. 

Typical methods to inform the choice of weights matrix include sequential specification testing, or model selection with reference to an information criterion \citep[e.g.,][]{zhang_spatial_2018}. These approaches have largely been focused on the problem of discerning a single best weights matrix from a set of mutually exclusive competitors. In contrast, there are many cases in which weights matrices manifest equally relevant, rather than competing, specifications and, in cases such as these, a model that includes multiple weights matrices may be preferable. This presents a more challenging model selection problem since prospective model specifications may be nested in one another, generating a large number of alternative models. In order to tackle this empirically important issue the current paper uses penalised estimation methods, which retain relevant weights matrix specifications, while at the same time, shrink the coefficients of irrelevant matrices to exactly zero. 

A related concern in models of this kind is unobserved heterogeneity. Intuitively, there are likely to be many common factors which are unobserved, and yet have an influence on the outcomes of cross-sectional units; for example exposure to common shocks or a common environment. The presence of common factors can make the identification of model parameters difficult in the event that these are correlated with covariates. The typical approach in dealing with unobserved heterogeneity is to transform the model in a way which purges the unobserved factors \citep[see, e.g.,][]{yu_quasi-maximum_2008,lee_estimation_nodate}. Nonetheless, a transform risks purging the very variation needed to identify network spillovers and therefore identification remains a delicate issue, with variation in the regressors, the structure of the weights matrices, and that of the unobserved heterogeneity, each having a part to play. An additional challenge in transforming the model is that prior knowledge on the nature of the unobserved heterogeneity is needed to specify a transform. Traditional examples of this include time, unit and group effect models in which case information on time, unit and group identities is used. Yet with a complex structure of cross-sectional interaction, it is desirable to go beyond these models and to allow for more general forms of
heterogeneity. The present paper models a factor structure in the error, which provides this flexibility since common factors may vary across time and have a fully heterogeneous effect on the cross-section. By way of principal component methods, a transform to
purge these factors is, in effect, estimated alongside model parameters, removing the reliance on prior knowledge to specify a transform. Taken together, multiple weights matrices, penalisation, and a factor structure error, provide a means of estimating various network spillovers which addresses some of the empirical concerns raised in models of cross-sectional interaction. Moreover, the properties of the estimator are studied in a framework where, although the true value of each coefficient is assumed to be fixed, the total number may be increasing with sample size. Such a regime better reflects the intuition that as sample size grows, so too is the amount of information available on a network likely to accumulate.

The present paper lies in the intersection of several literatures, including social and spatial network models, high-dimensional estimation, and models with factor structured errors. In the social
network literature, estimation and identification of network spillovers has been extensively discussed; e.g., \cite{lee_identification_2007}, \cite{bramoulle_identification_2009} and \cite{lee_specification_2010}. These papers each devote attention to the challenges which may arise in the presence of unobserved heterogeneity, in models where a time dimension is absent. Elsewhere, panel data models which combine interaction and factor structures in the error term have been considered; see, for example, \cite{shi_spatial_2017}, \cite{bai_dynamic_2017} and \cite{prucha}. In a likelihood framework, \cite{shi_spatial_2017} studies the estimation of a dynamic spatial model with interactive fixed effects and use a single weights matrix to represent dependencies between outcomes. \cite{bai_dynamic_2017} do similarly, though explicitly allowing for cross-sectional heteroskedasticity. The present paper also pursues a likelihood based estimation approach, and generalises these papers to allow for multiple weights matrices and the possibility that the number of such matrices may be increasing with sample size. \cite{prucha} consider estimation of a model with multiple potentially endogenous weights matrices alongside a factor structure in the error, by way of a method of moments estimator. The approach of \cite{shi_spatial_2017} is partly inspired by \cite{moon_linear_2015}, who derive the properties of an estimator using an eigenvalue perturbation approach. On the other hand, \cite{bai_dynamic_2017} more closely follow \cite{bai_panel_2009}, who derives results using first order conditions as a starting point for analysis. In terms of theory, this paper follows the latter approach, and proceeds from first order conditions in similar fashion to \cite{bai_panel_2009}. 

In the high-dimensional estimation literature, \cite{lu_shrinkage_2016} examine a model with interactive fixed effects and an increasing number of parameters, but without cross-sectional interaction. They make use of the adaptive Lasso penalty of \cite{zou_adaptive_2006} to induce sparsity amongst both estimated coefficients and factor loadings, assuming that many of these are redundant. Their procedure yields efficiency gains when compared to estimating the model with the number of factors overestimated. The present paper also uses the adaptive Lasso, which penalises the $\ell_1$ norm of the estimated parameter vector, encouraging sparsity amongst coefficient estimates. High-dimensional spatial models have also been studied elsewhere, such as \cite{Lam_estimation_2019}, who consider a model which allows for an increasing number of spatial weights matrices, and also use the adaptive Lasso as a penalty, though do not consider unobserved heterogeneity beyond standard fixed effect approaches. \cite{liu_doctor_2017} similarly uses penalised estimation in a cross-sectional model with many spatial weights matrices. \cite{gupta_inference_2015,gupta_pseudo_2018} consider estimation of a cross-sectional spatial model, with the number of weights matrices increasing with sample size, by using instrumental variables and quasi-maximum likelihood respectively. The authors carefully study the asymptotic behaviour of these estimators, but do not pursue penalised estimation nor discuss unobserved heterogeneity. 

Some recent works have also considered the case where the network is entirely unobserved, such as \cite{de_paula_recovering_nodate} and \cite{lewbel_2019}. This situation is especially relevant in the context of social interactions, where connections between individuals might be particularly hard to observe or to quantify. The approach taken in \cite{de_paula_recovering_nodate} involves estimating an entire weights matrix using observations on the same set of individuals across  multiple time periods. This can be seen as an extreme case of the current paper in which each weights matrix consists of a single nonzero element taking a value of one. \cite{lewbel_2019} takes a different perspective whereby multiple groups of individuals are observed, a special case of which is when each group consists of the same individuals observed in different time periods. In contrast, the focus of the present paper is where the network is partially observed, which in practice may be quite common. Moreover, establishing identification of the entire weights matrix once a factor term is introduced into the error may be a nontrivial matter.

\textbf{Outline}: The model of interest is introduced in Section \ref{ma}, alongside some basic assumptions and the estimation method. This is followed by asymptotic results in Section \ref{sec asy res}, and a discussion on implementation in Section \ref{imp}. In Section \ref{ill} finite sample performance is assessed be means of a small Monte Carlo study,
followed by an empirical application of the method to consider whether network spillovers are prevalent in determining growth rates across countries. Section \ref{concfl} concludes. Proofs of the main results can be found in Appendix \ref{mainres}. For further discussion, proofs of lemmas and additional
simulation output, see the Supplementary Material. 
  
\textbf{Notation}: Throughout the paper, all vectors and matrices are real.
For an $n \times 1$ vector $\bs{b}$ with elements $b_i$, $||\bs{b}||_1 \coloneqq
\sum^{n}_{i=1} |b_{i}|$, $||\bs{b}||_2 \coloneqq \sqrt{\sum^{n}_{i=1} b_{i}^2}$, $||%
\bs{b}||_{\infty} \coloneqq \max_{1\leq i \leq n}|b_i|$. Let $\bs{B}$ be an $%
n\times m$ matrix with elements $B_{ij}$. When $m =n$, and the eigenvalues
of $\bs{B}$ are real, they are denoted by $ \mu_n(\bs{B})
\leq \ldots \leq \mu_1(\bs{B})$. The following matrix norms
are those induced by their vector counterparts: $||\bs{B}||_1 \coloneqq \max_{1 \leq
j \leq m} \sum^{n}_{i=1} |B_{ij}|$ which is the maximum absolute column sum
of $\bs{B}$, $||\bs{B}||_2 \coloneqq \sqrt{\mu_{1}(\bs{B}^{\prime }\bs{B})}$, and $%
||\bs{B}||_{\infty} \coloneqq \max_{1\leq i \leq n} \sum^{m}_{j=1} |B_{ij}|$ which
is the maximum absolute row sum of $\bs{B}$. The Frobenius norm of $\bs{B}$
is denoted $||\bs{B}||_F \coloneqq \sqrt{\sum^{n}_{i=1} \sum^m_{j=1} B_{ij}^2} = 
\sqrt{\text{tr}(\bs{B}^{\prime }\bs{B})}$. Let $\bs{P}_{B} \coloneqq \bs{B}(\bs{B}%
^{\prime }\bs{B})^{+}\bs{B}^{\prime }$ and $\bs{M}_{{B}} \coloneqq \bs{I}_{n} - %
\bs{P}_{B}$, where $\bs{I}_m$ is the $m \times m$ identity matrix and the superscript $+$ denotes the Moore-Penrose generalised inverse. A sequence of $n \times n$ matrices $\bs{C}_n$ is said to be uniformly bounded in absolute row and column sums \textup{(UB)} if both the sequences $||\bs{C}_{n}||_{1}$ and $||\bs{C}_{n}||_{\infty}$ are bounded. Throughout $c$,
potentially indexed when there are many, is used to denote some
arbitrary positive constant which, unless stated otherwise, is assumed not to depend on sample size. Finally, `w.p.a.1' is used to indicate `with probability approaching 1'.

\section{Model and Estimation}

\label{ma}

\subsection{\label{sec model}Model}

The model studied in this paper supposes that, amongst $n$
cross-sectional units in time period $t=1,\ldots,T$, outcomes are generated
according to 
\begin{align}
\bs{y_{\textit{t}}} &= \sum^{Q_{nT}}_{q=1} \rho_q \bs{W}_{\kern -0.2em q} %
\bs{y_{\textit{t}}} + \sum^{K_{nT}}_{k=1} \beta_{k} \bs{x_{\textit{kt}}} + %
\bs{\eta}_t,  \label{A1}
\end{align}
where $\bs{y_{\textit{t}}},\bs{x_{\textit{kt}}}$ and $\bs{\eta}_t$ are $n
\times 1$ vectors of outcomes, covariates and error terms,
respectively, and $\bs{W}_{\kern -0.2em q}$ is an $n \times n$ weights matrix
specified in advance. Both the number $Q_{nT}$ of potentially relevant
weights matrices and the number $K_{nT}$ of potentially relevant regressors
can increase with sample size. The covariates may be subdivided into various types, such that 
\begin{align}
\sum^{K_{nT}}_{k=1} \beta_{k} \bs{x_{\textit{kt}}} 
= \sum^{K^*_{nT}}_{\kappa = 1} \delta_{\kappa} \bs{x}_{\kappa t}^*  + \phi_{1} \bs{y}_{t-1} +  \sum^{Q_{nT}}_{q = 1} \phi_{q+1} \bs{W}_{\kern -0.2em q} \bs{y}_{t - 1}. \label{A2}
\end{align}
The first ${K}^{*}_{nT}$ regressors may be either primitive exogenous covariates, or formed by the interaction of weights matrices and primitive exogenous covariates. It is assumed that there is at least one relevant exogenous covariate, i.e. this paper does not study the case of a pure network autoregression. Moreover, lagged outcomes and the interaction of these with weights matrices can provide additional covariates of the form $\bs{W}_{\kern -0.2em q} \bs{y}_{t-1}$. It may be that many of the parameters $\rho_q$, $\delta_\kappa$ and $\phi_q$ are truly zero since many of the covariates or weights matrix specifications may be irrelevant. Such restrictions need not be imposed a priori, since penalised estimation induces the estimates of these parameters to take values of exactly zero. 

The weights matrices $\bs{W}_{\kern -0.2em q}$ contain information about the
connections between the cross-sectional units, with larger
elements -- positive or negative -- measuring a stronger connection.
The literature often assumes that the weights matrices have positive
elements and are row normalised such that each of the rows of $%
\bs{W}_{\kern -0.2em q}$ sum to $1$. These assumptions lend products of the form $%
\bs{W}_{\kern -0.2em q} \bs{b}$ the
interpretation of a weighted average of the entries of a vector $\bs{b}$. While these two assumptions are not necessary in
this paper, the assumption that the weights matrices have zero diagonals,
which forbids self-links, is retained. The coefficients $\rho_q$ on $\bs{W}_{\kern -0.2em q} \bs{y}_t$ capture
endogenous spillovers; that is, the impact on the outcome of each unit, generated by the units that are neighbours according to the $q$-th weights matrix. Analogously, those $\delta_{\kappa}$ coefficients on covariates of the form $\bs{W}_{\kern -0.2em q} \bs{x}_{\kappa t}^*$ capture exogenous spillovers, also referred to as contextual effects in the social interaction literature. The coefficients $\phi_{q+1}$ on products $\bs{W}_{\kern -0.2em q} \bs{y}_{t -1}$ capture dynamic spillovers. Combined, the endogenous, exogenous and dynamic spillovers, allow model \eqref{A1} to quantify a breadth of different network spillovers.

It is assumed that the error term has a factor structure of the form 
\begin{align}
\bs{\eta}_t = \bs{\Lambda}\bs{f}_{\kern -0.1em t} + \bs{\varepsilon}_t,  \label{A2.5}
\end{align}
where $\bs{\Lambda}$ is an $n \times R$ matrix of time-invariant loadings, $\bs{f}_{\kern -0.1em t}$ is an $R \times 1$ vector of unit-invariant factors, and $\bs{\varepsilon}_t$ is an $n \times 1$ vector of idiosyncratic error terms. In addition, the rows of $\bs{\Lambda}$ are
denoted by $\bs{\lambda}_{i}$, for $i=1,\ldots,n$, and the factors are arranged in the $T \times
R$ matrix $\bs{F} \coloneqq (\bs{f}_{1},\ldots,\bs{f}_{T})^{\prime }$. Throughout, the number of factors $R$ is assumed to be a constant independent of sample size. Following a fixed effects approach, both factors and loadings are treated as
(nuisance) parameters in estimation. Thus, in the model, either is allowed to be arbitrarily correlated with covariates. The framework is very general; for instance $\bs{f}_{\kern -0.1em t}$ could be aggregate shocks affecting the entire network at time $t$, with a heterogenous effect on each individual. Moreover, this factor structure
nests more traditional fixed effect models as special cases.

It is worth stressing that unobserved heterogeneity may arise from various sources. 
Consider, as a simple example, a model with a single exogenous regressor and no endogenous spillovers, i.e., 
\begin{align}
\bs{y_{\textit{t}}} &= \beta^*\bs{x}^*_{\mathit{t}} + \sum^{Q_{nT}}_{q=1}
\alpha_q \bs{W}_{\kern -0.2em q} \bs{x}^*_{\mathit{t}} + \delta %
\bs{W_{\textit{L}}} \bs{x}^*_{\mathit{t}} + \bs{\varepsilon}_t,  \label{A3}
\end{align}
with $\bs{W}_{\kern -0.2em q}$ being the $q$-th observed weights matrix, and $%
\beta^*,\alpha_{q},\delta$ being scalars. Suppose that $\bs{W_{\textit{L}}}$ is unobserved and is either low rank or
well approximated by a low rank matrix. This may represent, for example, low rank measurement error in some $\bs{W}_{\kern -0.2em q}$, or unobserved connections between cross-sectional units arising due to network sampling; see, for instance, \cite{yike}. Defining $\bs{\Lambda}%
^* \bs{f}_{\kern -0.1em t}^* \coloneqq \delta \bs{W_{\textit{L}}} \bs{x}^*_{\mathit{t}}$, it
is clear that \eqref{A3} is nested in model \eqref{A1} and highlights that the decomposition of the unobserved term into factors $\bs{\Lambda}^*$ and loadings $\bs{f}_{\kern -0.1em t}^*$ is arbitrary; it is the low rank restriction on $\delta \bs{W_{\textit{L}}} \bs{x}^*_{\mathit{t}}$ that allows this term to be distinguished and controlled for.

Going forward, it is convenient to introduce some new notation.  The subscript $nT$ used previously is suppressed from $Q_{nT}$, $K_{nT}$, $K^{*}_{nT}$, and the following parameter vectors and covariate matrices are
defined: $\bs{\rho} \coloneqq (\rho_1,\ldots,\rho_{Q})^{\prime }$, $\bs{\delta} \coloneqq (\delta_1,\ldots,\delta_{K^{*}})'$, $\bs{\phi} \coloneqq (\phi_1,\ldots,\phi_{Q+1})'$,  $\bs{\beta} \coloneqq(\beta_1,\ldots,\beta_K)^{\prime }\coloneqq (\bs{\delta}', \bs{\phi}')'$, $\bs{\theta} \coloneqq (\bs{\rho}^{\prime },%
\bs{\beta}^{\prime })^{\prime }$, and $\bs{X_\textit{t}}\ \coloneqq (\bs{X_\textit{t}^*}, \bs{y}_{t-1}, \bs{W_\text{1}}\bs{y}_{t-1},\ldots,\bs{W}_{\kern -0.2em Q}\bs{y}_{t-1})$, where $\bs{X_\textit{t}^*} \coloneqq (\bs{x}_{1t}^*,\ldots,\bs{x}%
_{K^* t}^*)$, and $\bs{S}(\bs{\rho}) \coloneqq %
\bs{I}_n - \sum^{Q}_{q=1} \rho_q \bs{W}_{\kern -0.2em q}$. Given these, model \eqref{A1} can be restated more succinctly
as 
\begin{align}
\bs{S}(\bs{\rho}) \bs{y_{\textit{t}}} = \bs{X}_t \bs{\beta} + \bs{\Lambda}\bs{f}%
_t + \bs{\varepsilon}_t.  \label{A3.1}
\end{align}
Throughout, the superscript $0$ is used to distinguish the true values of the factors, loadings, and parameters, as well as the true numbers of these, and the framework is one in which $n$ and $T$ diverge simultaneously. The total number of parameters in the vector $\bs{\theta}$ is $P \coloneqq Q + K$, of which only $%
P^0$ are truly nonzero. In fact, one might often expect that
the vector $\bs{\theta}$ is sparse in the sense that many of its
components are zero, particularly in cases with a large number of weights matrices and covariates. Accordingly $\bs{\theta}$ may be reordered as $\bs{\vartheta} \coloneqq (\bs{\theta}_{(1)}',\bs{\theta}_{(2)}')'$, where $\bs{\theta}_{(1)}$ is the $P^0\times 1$ vector of nonzero parameters, and $\bs{\theta}_{(2)}^0=\bs{0}_{ (P - P^0) \times 1 }$. Sparsity, however, is not necessary and indeed the results of this paper equally allow for the possibility that all of the weights matrices and covariates may be relevant. 
The $n \times T$ data matrix for the $\kappa$-th exogenous covariate is 
denoted $\bs{\mathcal{X}}_\kappa^* \coloneqq (\bs{x}_{\kappa 1}^*,\ldots,\bs{x}_{\kappa T}^*)$ for $\kappa = 1,\ldots,K^*$, and the $n \times T$ data matrix for the lagged outcomes is denoted $\bs{\mathcal{Y}}_{-1} \coloneqq (\bs{y}_{t-1},\ldots,\bs{y}_{t-T})$. The data matrix for the generic $k$-th
covariate of any type, $\bs{\mathcal{X}}_\kappa^*$, $\bs{\mathcal{Y}}_{-1}$ or $\bs{W}_{\kern -0.2em q}%
\bs{\mathcal{Y}}_{-1}$, is denoted $\bs{\mathcal{X}}_k \coloneqq (\bs{x}_{k
	1},\ldots,\bs{x}_{k T})$, for $k = 1,\ldots,K$. Also, $\bs{A}(\bs{\rho},\bs{\phi}) \coloneqq \bs{S}^{-1}(\bs{\rho})(\phi_1 \bs{I}_n + \sum^Q_{q=1} \phi_{q+1} \bs{W}_{\kern -0.2em q})$, $\bs{A} \coloneqq  \bs{A}(\bs{\rho}^0,\bs{\phi}^0)$, $\bs{S} \coloneqq \bs{S}(%
\bs{\rho}^0)$, $\bs{G}_{q}(\bs{\rho}) \coloneqq \bs{W}_{\kern -0.2em q} \bs{S}^{-1}(\bs{\rho})$, and $\bs{G}_q \coloneqq \bs{G}_q(\bs{\rho}^0)$.

\subsection{Assumptions}
The first set of assumptions concerns the idiosyncratic error term ${\varepsilon_{it}}$. 
\begin{Assumption}\label{ASS1} \color{white}.\color{black}\
\begin{enumerate}[label = 1.\arabic*]
\item The errors ${\varepsilon}_{it}$ are identically and independently distributed over $i$ and $t$ with $\bs{\mathbb{E}}[\varepsilon_{it}] = 0$, $\bs{\mathbb{E}}[\varepsilon_{it}^2] = \sigma^2_0 \geq  c  > 0$, and fourth moments uniformly bounded over $i$ and $t$. \label{AS:1.1} 
\item The errors ${\varepsilon}_{it}$ are independent of the elements of the matrices $\bs{\Lambda}^0$, $\bs{F}^0$, and $\bs{\mathcal{X}}_{\kappa}^*$, for $\kappa = 1,\ldots,K^*$. \label{AS:1.2} 
\end{enumerate}
\end{Assumption}

These assumptions have been employed across various papers. Cross-sectional
homoskedasticity and independence is commonly assumed, though this can be relaxed by
estimation of a more general $n \times n$ covariance matrix $\bs{\Sigma}^0$,
at the expense of additional parameters; see for example \cite%
{bai_inferences_2017} and \cite{bai_dynamic_2017}.
Additional structure in the error term could also be considered as is
commonplace throughout the spatial econometrics literature. Yet since the
factor structure provides a mechanism for capturing such correlation,
Assumption \ref{AS:1.1} assumes $\bs{\Sigma}^0 = \sigma^2_0 \bs{I}_n$.
Differing assumptions concerning the relationship between the errors, the factors, and the loadings
appear across the literature; these are comprehensively surveyed by \cite{hsiao_panel_2018}. Assumption \ref{AS:1.2} imposes independence
of the factors and the loadings from the error term as in \cite{bai_panel_2009}.

Some additional assumptions are required regarding the other components of the model. Let $|\cdot|$ denote the entrywise absolute value of a vector or matrix, $\bs{\Theta}$ denote the parameter space for $\bs{\theta}$, and $\bs{\Theta}_{{\rho}}$ and $\bs{\Theta}_{{\phi}}$ denote the parameter spaces for $\bs{\rho}$ and $\bs{\phi},$ respectively. Since $\bs{\Theta}$, $\bs{\Theta}_{{\rho}}$ and $\bs{\Theta}_{{\phi}}$ depend on $n$ and $T$, in the following it is understood that any assumptions which relate to these parameter spaces are to hold for any $(n,T)$. 
\begin{Assumption}\label{ASS2} \color{white}.\color{black}	
\begin{enumerate}[label = 2.\arabic*]
\item The parameter vector $\bs{\theta}^0$ is in the interior of
$\bs{\Theta}$, with $\bs{\Theta}$ being a compact subset of $\bs{\mathbb{R}}^{P}$. \label{AS:2.2}
\item The weights matrices $\bs{W}_{\kern -0.2em q}$ are nonstochastic and \textup{UB} uniformly over $q$.    \label{AS:2.3}
\item For all $\bs{\rho} \in \bs{\Theta}_{\rho}$ and $\bs{\phi} \in \bs{\Theta}_{\phi}$, $\bs{S}(\bs{\rho})$ is invertible,  $\bs{S}(\bs{\rho}), \bs{S}^{-1}(\bs{\rho})$ and $\sum^{\infty}_{h=1} |\bs{A}^h(\bs{\rho},\bs{\phi})|$ are \textup{UB},  $||\bs{A}(\bs{\rho},\bs{\phi})||_2 < 1 - c$ for some $c>0$, and $\liminf_{n,T \rightarrow \infty} \inf_{\bs{\rho} \in \bs{\Theta}_{{\rho}}}  \textup{det} (\bs{S}(\bs{\rho})) \neq 0$. \label{AS:2.4} \vspace{-0.5cm}
\item The elements of the matrices $\bs{\mathcal{X}}^*_{\kappa}$ have fourth moments uniformly bounded over $i,t$ and $\kappa$, and elements of the matrix $\sum^{K}_{k=1} \beta^0_k \bs{\mathcal{X}}_{k}$ have fourth moments uniformly bounded over $i,t$ and $K$. \label{AS:2.5} 
\item The true number of factors ${R}^0$ is constant. \label{AS:2.7}
\item The elements of the matrices $\bs{F}^0$ and $\bs{\Lambda}^0$ have eighth moments uniformly bounded over $i$ and $t$. \label{AS:2.8}
\end{enumerate}
\end{Assumption} 
Assumption \ref{AS:2.2} considers a sequence of compact parameter
spaces over which to maximise the objective function. The condition in Assumption 
\ref{AS:2.3} that the weights matrices are UB is standard and serves to limit
interactions to a manageable degree. Here, uniform boundedness over $q$ is also required, due to the possibility that the number of weights matrices increases with sample size. Assumption \ref{AS:2.4} ensures that the model
admits a reduced form, and the dynamic process in stationary. Restrictions on the parameter space of $\bs{\rho}$
which ensure that $\bs{S}(\bs{\rho})$ is invertible have been discussed elsewhere in
the literature, particularly in the case $Q = 1$. A general condition
sufficient for the invertibility of $\bs{S}(\bs{\rho})$ is $||\sum^Q_{q=1}
\rho_q \bs{W}_{\kern -0.2em q}|| < 1$ for some matrix norm $||\cdot||$, though with $Q > 1$ more
informative conditions can be difficult to obtain outside of exceptional
cases.\footnote{One such case is where the matrices $\bs{W_{\textup{1}}},\ldots,%
\bs{W}_{\kern -0.2em Q}$ are simultaneously diagonalisable, for example where they consist of powers of a single weights matrix. Another example is where the
weights matrices consist of nonoverlapping blocks.}
However, as noted by \cite{gupta_pseudo_2018}, even when it is possible to
characterise inadmissible values of $\bs{\rho}$ and exclude these, the
resulting parameter space is unlikely to be compact. It is therefore
commonplace in the literature to restrict attention to a region around the
origin in which $\bs{S}(\bs{\rho})$ can be guaranteed to be invertible. This
is where $\sum^Q_{q=1} |\rho_q| < (\max_{1 \leq q \leq Q} ||%
\bs{W}_{\kern -0.2em q}||)^{-1} $.\footnote{This inequality is obtained from the condition $||\sum^Q_{q=1}
	\rho_q \bs{W}_{\kern -0.2em q}|| < 1$ and the fact that $||\sum^Q_{q=1}
	\rho_q \bs{W}_{\kern -0.2em q}|| \leq \sum^Q_{q=1}
	|\rho_q| \max_{1 \leq q \leq Q} ||\bs{W}_{\kern -0.2em q}||$ for any matrix norm $||\cdot||$.} Yet while the set of $\bs{\rho}$ which
satisfy this is bounded, it is also open. Therefore to ensure the existence
of a maximiser over this space, a closed subset can be considered such that $%
\sum^Q_{q=1} |\rho_q| \leq (1 - \tau)(\max_{1 \leq q \leq Q} ||%
\bs{W}_{\kern -0.2em q}|| )^{-1} $, with $\tau \in (0,1)$. Row normalisation
of the matrices $\bs{W}_{\kern -0.2em q}$ further
simplifies this condition since it implies $\max_{1 \leq q \leq Q} ||%
\bs{W}_{\kern -0.2em q}||_{\infty} = 1$. Model \eqref{A3.1} can be rewritten as $\bs{y}_t = \bs{A}\bs{y}_{t-1} + \bs{S}^{-1}(\bs{X}^*_t \bs{\delta} + \bs{\Lambda}\bs{f}_{\kern -0.1em t} + \bs{\varepsilon}_t)$, or, after recursive substitution, $\bs{y}_t = \sum^{\infty}_{h=0} \bs{A}^h \bs{S}^{-1} (\bs{X}^*_{t-h} \bs{\delta} + \bs{\Lambda} \bs{f}_{t-h} + \bs{\varepsilon}_{t-h})$; Assumption \ref{AS:2.4} guarantees that this series converges. Further discussion of parameter restrictions ensuring convergence of this series can be found in \cite{SpatialPanelDataModels} and \cite{shi_spatial_2017}.\footnote{For example, where the
	weights matrices consist of nonoverlapping blocks, $\sum^Q_{q=1}|\rho_q| + \sum^{Q+1}_{q=1}|\phi_q| < 1 - c$ is sufficient for $||\bs{A}(\bs{\rho},\bs{\phi})||_2 < 1 - c$.} The first part of Assumption \ref{AS:2.5} ensures that $%
||\bs{\mathcal{X}}_k^*||_F = O_P(\sqrt{nT})$, for $k=1,\ldots,K$. For the second part, notice that $\bs{G}_q \bs{{X}}_t \bs{\beta}^0$ can be used as
an instrument in the estimation of ${\rho}_q$.\footnote{Observing that $\bs{S}^{-1} = \bs{I}_n + \sum^{Q}_{q=1} \rho_q^0 \bs{G}_q$, then $
\bs{y_{\textit{t}}} 
= \bs{X}_t \bs{\beta}^0 + \sum^Q_{q=1} \rho^0_q \bs{G}_q \bs{X}_t \bs{\beta}^0  + \bs{S}^{-1} \bs{\Lambda}^0\bs{f}_{\kern -0.1em t}^0 + \bs{S}^{-1} \bs{\varepsilon}_t$, which makes the role of $\bs{G}_q \bs{{X}}_t \bs{\beta}^0$ as an instrument for $\bs{W}_{\kern -0.2em q}\bs{y}_t$ transparent.} With a diverging number of parameters, the second part of Assumption \ref{AS:2.5} assures that for these 
instruments $||\sum^{K}_{k=1} \beta^0_k 
\bs{G}_q \bs{\mathcal{X}}_{k}||_F = O_P(\sqrt{nT})$. An alternative condition sufficient for this is $||\bs{\beta}^0||_1 < c$, which follows by H\"{o}lder's
inequality or, alternatively, Assumption \ref{AS:2.5} could be replaced by one
restricting the growth of $K^0$ and $n,T$. Assumption \ref{AS:2.7} is common throughout the literature, but could be relaxed at the expense of
slower rates of convergence. Several differing assumptions concerning the
moments of the factors and the loadings appear in the literature. Given the possible presence of lagged outcomes as covariates, Assumption \ref{AS:2.8} serves the same purpose as Assumption 5(vi) in \cite{moon_dynamic_nodate}, and ensures that $y_{it}$ has uniformly bounded fourth moments.

\subsection{Objective Function}\label{em}

The estimation strategy employed in this paper is penalised quasi-maximum
likelihood (PQML), using the multivariate standard normal distribution for
the error term, i.e., ${\varepsilon }_{it}\overset{\text{iid}}{\sim }\ 
\mathcal{N}(0,\sigma _{0}^{2})$, and following a fixed effects approach. Maximum likelihood estimation is a standard in the literature for models of this type, since the simultaneity in the determination of outcomes generates an endogeneity problem which results in least squares estimates being biased. The parameter of interest is $\boldsymbol{\theta }$, whereas $\boldsymbol{\Lambda },\boldsymbol{F},\sigma ^{2}$ are treated as nuisance parameters. Since fixing $\boldsymbol{\theta }$ results in a pure factor model (and $
\boldsymbol{\Lambda },\boldsymbol{F},\sigma ^{2}$ are not penalised), the
estimators of $\boldsymbol{\Lambda }$ and $\boldsymbol{F}$ for fixed $%
\boldsymbol{\theta }$ are a solution to a standard principal component
problem 
\citep[see,
e.g.,][]{bai_panel_2009}. In this subsection $R$ is fixed such that $R \geq R^0$; this is discussed in greater detail in Section \ref{cons}. With $R$ fixed, the average (quasi) log-likelihood is  
\begin{IEEEeqnarray}{rCl}
\mathcal{L}(\bs{\theta},\bs{\Lambda},\bs{F},\sigma ^{2}) &\coloneqq & -\frac{1}{2}\log
(2\pi )+\frac{1}{n}\log (\det (\bs{S}(\bs{\rho})))-\frac{1}{2}\log (\sigma
^{2})  \nonumber \\
&& -\frac{1}{2\sigma ^{2}}\frac{1}{nT}\sum_{t=1}^{T}(\bs{S}(\bs{\rho})%
\bs{y_\textit{t}}-\bs{X_{\textit{t}}}\bs{\beta}-\bs{\Lambda}\bs{f_\textit{t}}%
)^{\prime }(\bs{S}(\bs{\rho})\bs{y_\textit{t}}-\bs{X_{\textit{t}}}\bs{\beta}-%
\bs{\Lambda}\bs{f_\textit{t}})  \label{A4}
\end{IEEEeqnarray}%
and its penalised counterpart is 
\begin{equation}
\mathcal{Q}(\bs{\theta},\bs{\Lambda},\bs{F},\sigma ^{2}) \coloneqq \mathcal{L}(%
\bs{\theta},\bs{\Lambda},\bs{F},\sigma ^{2})-\varrho(\bs{\theta},\bs{\gamma},{\zeta}),  \label{A5}
\end{equation}%
where $\varrho(\bs{\theta},\bs{\gamma},{\zeta})$ is a penalty function and $\bs{\gamma}$, ${\zeta}$ are regularisation
parameters. The specific form of penalty function is introduced in Section \ref{Penalty}, and the choice of regularisation parameters is discussed in Section \ref{penparam}, however for the moment these are both also taken to be fixed alongside the number of factors. Concentrating out $\sigma ^{2}$, as well as the factors, and
dropping the constant in \eqref{A5} yields the concentrated expression 
\begin{align}
\mathcal{Q}(\bs{\theta},\bs{\Lambda})\coloneqq\frac{1}{n}\log (\det (\bs{S}(\bs{\rho}%
)))-\frac{1}{2}\log \left( \hat{\sigma}^{2}(\bs{\theta},\bs{\Lambda})\right)
-\varrho(\bs{\theta},\bs{\gamma},{\zeta}), \label{A8}
\end{align}
where $\hat{\sigma}^{2}(\bs{\theta},\bs{\Lambda}) \coloneqq \frac{1}{nT}\sum_{t=1}^{T} \bs{e}_t^{\prime}
\bs{M}_{\bs{\Lambda}} \bs{e}_t$ and $\bs{e}_{t}\coloneqq \bs{S}(\bs{\rho})\bs{y_\textit{t}}-\bs{X_{\textit{t}}}%
\bs{\beta}$. Hereafter the terms likelihood and log-likelihood are
used synonymously. In order to maximise \eqref{A8} with respect to $\bs{\Lambda}$, note that
\begin{IEEEeqnarray}{rCl}
\min_{\bs{\Lambda} \in \mathbb{R}^{n \times R}} \frac{1}{nT}\sum_{t=1}^{T} \bs{e}_t^{\prime}
&\bs{M}_{\bs{\Lambda}}& \bs{e}_t
=  \frac{1}{nT} \sum_{t=1}^{T} \bs{e}_t^{\prime} \bs{e}_t   - \max_{\bs{\Lambda} \in \mathbb{R}^{n \times R}} \frac{1}{nT}\sum_{t=1}^{T} \bs{e}_t^{\prime}
\bs{P}_{\bs{\Lambda}} \bs{e}_t \notag \\
&=& \text{tr}\left( \frac{1}{nT} \sum_{t=1}^{T}\bs{e}_t \bs{e}_t^{\prime}  \right) -  \max_{\bs{\mathcal{V}}_{\bs{\Lambda}}  \in \mathbb{R}^{n \times R} : \bs{\mathcal{V}}_{\bs{\Lambda}} '\bs{\mathcal{V}}_{\bs{\Lambda}} = \bs{I}_R} \text{tr} \left(\frac{1}{nT}\sum_{t=1}^{T} \bs{\mathcal{V}}_{\bs{\Lambda}}  ' \bs{e}_t\bs{e}_t^{\prime}
\bs{\mathcal{V}}_{\bs{\Lambda}}   \right) \IEEEeqnarraynumspace \notag \\
&=& \text{tr}\left( \frac{1}{nT} \sum_{t=1}^{T}\bs{e}_t \bs{e}_t^{\prime}  \right) -  \sum^R_{r = 1} \mu_r \left( \frac{1}{nT}\sum_{t=1}^{T} \bs{e}_t\bs{e}_t^{\prime}  \right), \label{eeee}
\end{IEEEeqnarray}
where the second line follows from the fact that any orthogonal projector $\bs{P}_{\bs{B}}$ can be written as $\bs{\mathcal{V}}_{\bs{B}} \bs{\mathcal{V}}_{\bs{B}}'$, with the columns of $\bs{\mathcal{V}}_{\bs{B}}$ forming an orthonormal basis for the column space of $\bs{B}$, and the third line follows from a standard result \citep[e.g.,][Corollary 4.3.39]{Horn}.\footnote{For example, by the QR decomposition, $\bs{B} = \bs{\mathcal{V}}_{\bs{B}} \bs{\mathscr{R}}$ with $ \bs{\mathcal{V}}_{\bs{B}}  \in \mathbb{R}^{n \times m}$ having orthonormal columns and $\bs{\mathscr{R}} \in \mathbb{R}^{m \times m}$ being upper triangular. Since $\bs{B}$ has full column rank $\bs{\mathscr{R}}$ is invertible \citep[e.g.,][Theorem 2.1.14]{Horn} and therefore $\bs{P}_{\bs{B}} \coloneqq \bs{B}(\bs{B}'\bs{B})^{-1}\bs{B} =\bs{\mathcal{V}}_{\bs{B}} \bs{\mathcal{V}}_{\bs{B}} ' $.}
Hence, \eqref{eeee} can be used to
concentrate out $\bs{\Lambda}$ in \eqref{A8}, whereby the PQML estimator of $%
\bs{\theta}^{0}$ is characterised as 
\begin{align}
\bs{\hat{\theta}} 
\coloneqq
\argmax_{\bs{\theta} \in \bs{\Theta}}\mathcal{Q}(\bs{\theta}),
\end{align}
where 
\begin{align}
\mathcal{Q}(\bs{\theta}) \coloneqq \frac{1}{n}\log (\det (\bs{S}(\bs{\rho})))-\frac{1}{%
2}\log \left( \sum_{i=R+1}^{n}\mu _{i}\left( \frac{1}{nT}\sum_{t=1}^{T}\bs{e}%
_{t}\bs{e}_{t}^{\prime }\right) \right) -\varrho(\bs{\theta},\bs{\gamma},{\zeta}). \label{objfun}
\end{align}
It is worth highlighting that both the factors and the loadings have been concentrated out without imposing any of the normalisations typically encountered in the wider factor literature. This is due to the treatment of both the factors and the loadings as nuisance parameters, in which case only the space spanned by the loadings implicitly features in the objective function \eqref{objfun}. It would, of course, be possible to consider estimators of the factors and the loadings, however the same fundamental indeterminacy issue would arise in separating these as is encountered elsewhere in the factor literature, and therefore some normalisations would typically be required in order to do this. It should also be pointed out that neither the concentrated likelihood nor the penalised objective function $\mathcal{Q}(\bs{\theta})$ are concave in $\bs{\theta}$. Although subsequent sections establish the desirable asymptotic properties of global maximisers of these objective functions, it is nonetheless the case that local maximisers which do not possess these properties may indeed exist. 

\subsection{Penalty}\label{Penalty}
The present paper adopts the adaptive Lasso, which induces sparsity in parameter estimation
by augmenting an objective function with a constraint on the $\ell _{1}$
norm of the estimated parameter vector. A desirable feature of this method of penalisation is that it can achieve the oracle property; that is, perform
consistent variable selection and, at the same time, possess an optimal rate of convergence. This is done by using an initial consistent estimator of the parameters to
weight the penalty. The cost of this is the need to find an initial consistent
estimator, which can be difficult in settings where the number of
parameters is greater than the number of observations ($nT<P$ in the present case).  This complication is not considered in this paper and
attention is restricted to the $nT>P$ setting. Explicitly, the penalty function employed in this paper has the additive form
\begin{align}
\varrho(\bs{\theta},\bs{\gamma},{\zeta}) 
\coloneqq 
\gamma_{\rho} \sum^{Q}_{q=1} \omega_{q} |\rho_{q}|
+
\gamma_{\beta} \sum^{K}_{k=1} \omega_{Q+k} |\beta_{k}|,
\label{penf}
\end{align}
where $\omega_{p} \coloneqq |\theta_{p}^{\dagger}|^{-\zeta}$, with $\theta_{p}^{\dagger}$ being an initial consistent estimate of the $p$-th parameter, and $\bs{\gamma} \coloneqq (\gamma_{\rho},\gamma_{\beta})'$ and $\zeta$ are regularisation parameters.\footnote{If $\theta_{p}^{\dagger} = 0$ then $\omega_p$ is set equal to $\infty$.} The parameter $\zeta$ is a positive constant and is used to adjust the weight of penalisation according the rate of consistency of the initial estimator. Combined, $\zeta$ and $\theta_{p}^{\dagger}$ generate bespoke weights $\omega_p$ for each parameter that will increase for truly zero coefficients and tend to a constant for truly nonzero coefficients. The other penalty parameters $\gamma_{\rho}$ and $\gamma_{\beta}$ are positive sequences which tend towards zero as $n$ and $T$ increase. The form of the penalty term in \eqref{penf} allows the penalty parameters $\gamma_{\rho}$ and $\gamma_{\beta}$ to differ across the two types of parameter, $\rho_q$ and $\beta_k$. In general the penalty term can be easily modified to allow for a greater or lesser degree of heterogeneity, as applications dictate. 

Let $\underline{\theta}^0$ and $\bar{\theta}^0$ denote, respectively, the minimum and maximum element of $|\bs{\theta}_{(1)}^0|$. Note that both $\underline{\theta}^0$ and $\bar{\theta}^0$ can vary with sample size due to the increasing dimension of $\bs{\theta}_{(1)}^0$. The following are assumed.
\begin{Assumption}\label{ASS3} \color{white}.\color{black}\
	\begin{enumerate}[label = 3.\arabic*]
		\item $0 < c_1 \leq \underline{\theta}^0 \leq \bar{\theta}^0 \leq c_2 < \infty$. \label{AS:4.0}
		\item $\max\{\gamma_{\rho},\gamma_{\beta}\}\min\{n,T\} = O(1)$. \label{AS:4.1}
		\item $||\bs{\theta}^{\dagger} - \bs{\theta}^0||_2 = O_P(r_{nT})$, for some sequence $r_{nT} \rightarrow 0$ as $n,T \rightarrow \infty$. \label{AS:4.2}
	\end{enumerate}
\end{Assumption}
In this paper it is assumed that, while the dimension of $\bs{\theta}^0$ may be increasing with sample size, the value of each element is fixed.\footnote{More precisely, it is assumed that $\theta^0_p$ does not depend on $n,T$, for any $n,T$ such that $\theta_p$ enters the model.} Nonetheless, this does not rule out either the minimum or maximum (in absolute value) nonzero elements in $\bs{\theta}^0$ becoming arbitrarily small or large as its dimension increases, and therefore Assumption \ref{AS:4.0} imposes that the nonzero elements in $\bs{\theta}^0$ are uniformly bounded away from zero and from infinity. Assumption \ref{AS:4.1} requires the penalty parameters $\gamma_{\rho}$ and $\gamma_{\beta}$ to converge to zero sufficiently fast that they do not adversely impact the rate of consistency of the estimator. Assumption \ref{AS:4.2} requires consistency of the initial estimator $\bs{\theta}^{\dagger}$ at some rate $r_{nT}$. If the speed at which $r_{nT} \rightarrow 0$ is especially slow, then $\zeta$ can be adjusted to compensate for this. In the following it is shown that the unpenalised likelihood can be used to produce a initial consistent estimator, though other estimation procedures might equally be considered.

In principle it would also be possible to obtain several of the results in this paper under a `moving parameter' framework, where the values of the nonzero elements in $\bs{\theta}^0$ might vary with sample size; in particular, where some may converge to zero asymptotically. However, the rate at which they could be allowed do so would need to be sufficiently slow that a choice of $\gamma_{\rho}$ and $\gamma_{\beta}$ could still be made to ensure the consistency and model selection consistency of the procedure.
Moreover, in Section \ref{asydis} the assumption that the value of nonzero elements in $\bs{\theta}^0$ are fixed is important for the validity of the asymptotic distribution derived in that section. Therefore, this assumption is maintained throughout this paper. 

\color{black}

\section{Asymptotic Results}\label{sec asy res} 
\subsection{Consistency}\label{cons} 
Mirroring \cite{bai_panel_2009}, in this section a preliminary consistency result is established which will be improved upon later. Yet, before proceeding, it is worth providing a few remarks on the identification of model parameters. In the standard consistency argument for an extremum estimator, the essence of
the idea is to show that \textit{\textquotedblleft the limit of the maximum $\bs{\hat{\theta}}$ should be the maximum of the limit"}, with the latter being unique \citep[][p.\color{white}.\color{black}2120]{Newey}. In that argument the role that identification plays is transparent, and with identification established, uniform
convergence of the sample objective function to the limiting objective function often then appeals to a uniform law of large numbers, and consistency follows thereafter. Yet in models where the number of
parameters, nuisance or otherwise, depends on the sample size, there is no fixed population distribution from which a sample is drawn, and therefore uniform convergence must be considered more carefully. In cases such as these, consistency is often shown directly, forgoing an explicit identification result. For these same reasons this paper also proceeds directly to consistency, with further discussion on identification being available in Appendix B of the Supplementary Material.

Before formulating the next assumption, it is necessary to introduce some additional notation.  Define the $n \times P$ matrix of instruments $\bs{{Z}_{\textit{t}}} \coloneqq (\bs{G}_{1} \bs{{X}}%
_t \bs{\beta}^0,\ldots,\bs{G}_{Q} \bs{{X}}_t \bs{\beta}^0,\bs{X}_{t})$. The $n \times T$ data matrix for
the instrument associated with some ${\rho}_q$ is $\sum^{K}_{k=1} \beta^0_k %
\bs{G}_q \bs{\mathcal{X}}_k$. The generic $n\times T$ data matrix of
either type, $\bs{\mathcal{X}}_{k}$ or $\sum^{K}_{k=1} \beta^0_k \bs{G_q} %
\bs{\mathcal{X}}_k$, is denoted $\bs{\mathscr{Z}}_p \coloneqq (\bs{z}_{p1},\ldots,\bs{z}%
_{pT})$, where $\bs{z}_{pt}$ is the $p$-th column of $\bs{Z}_t$, for $p =
1,\ldots,P$. Finally, let
$\bs{\mathcal{H}}_1(\bs{\Lambda},\bs{F}) \coloneqq \frac{1}{nT} \bs{\mathcal{Z}}^{\prime }(\bs{M}_{\bs{F}} \otimes \bs{M}_{\bs{{\Lambda}}}) \bs{\mathcal{Z}}$ and $\bs{\mathcal{H}}_2 \coloneqq \frac{1}{nT} \bs{\mathcal{Z}}^{\prime }\bs{\mathcal{Z}}$, 
where $\bs{\mathcal{Z}} \coloneqq (\bs{Z}_1',\ldots,\bs{Z}_{T}')'$ is a $nT \times P$ matrix.
\begin{Assumption}\label{ASS4} \color{white}.\color{black}\
	\begin{enumerate}[label = 4.\arabic*]	
\item $R \geq R^0$. \label{AS:6.4}		
\item $\inf_{\bs{{\Lambda}} \in \mathbb{R}^{n \times R}, \bs{F} \in \mathbb{R}^{T \times R^0}} \mu_{P}(\bs{\mathcal{H}}_1( \bs{\Lambda},\bs{F}) ) \geq c_1 > 0$ w.p.a.1 as $n,T \rightarrow \infty$. \label{new1}
\item $\mu_{1}(\bs{\mathcal{H}}_2) \leq c_2 < \infty$ w.p.a.1 as $n,T \rightarrow \infty$. \label{new2} 	
\item $\frac{P}{\min\{n,T\}} \rightarrow 0$. \label{AS:6.3}	
	\end{enumerate}
\end{Assumption} 
Assumption \ref{AS:6.4} allows for the true number of factors $R^0$ to be unknown, as long as the number of factors $R$ used in estimation is no less than $R^0$; see \cite{moon_linear_2015}. Assumption \ref{new1} demands a certain level of variation in sample data
after projecting out arbitrary factors and factor loadings. This condition can intuitively be understood by considering the particular case of individual or time effects, in which case the projections perform between individual and between time period differences to the data. It is also worth noting that Assumptions \ref{new1} and \ref{new2} imply that, w.p.a.1,
\begin{align}
\sup_{\bs{{\Lambda}} \in \mathbb{R}^{n \times R}, \bs{F} \in \mathbb{R}^{T \times R^0}} \mu_{1} \left(  \bs{\mathcal{H}}_1( \bs{\Lambda},\bs{F})  \right) \leq c_2 < \infty  \label{A1.21.1}
\end{align}
and
\begin{align}
\mu_{P} \left( \bs{\mathcal{H}}_2 \right) \geq c_1 >0,  \label{A1.21}
\end{align}
which ensures both $\bs{\mathcal{H}}_1$ and $\bs{\mathcal{H}}_2$ are well defined asymptotically (see Appendix C in the Supplementary Material for details). Assumption \ref{AS:6.3} requires that the
number of parameters does not grow too fast in relation to $n$ and $T$. This
is necessary since consistency is stated in terms of the $\ell_2$ norm of a
vector with increasing dimension. Recall that $\bs{\hat{\theta}}$ denotes the maximiser of the penalised likelihood function and let $\bs{\tilde{\theta}}$ denote the maximiser of the unpenalised likelihood function.
\begin{proposition}[Consistency]\label{Prop2}
	Under Assumptions \ref{ASS1}--\ref{ASS4}, $||\bs{\tilde{\theta}} - \bs{\theta}^0||_2 = O_P\left( a_{nT} \right) $ and $||\bs{\hat{\theta}} - \bs{\theta}^0 ||_2  = O_P\left( a_{nT} \right)$, where $a_{nT} \coloneqq \sqrt{\frac{P}{\min\{n,T\}}}$. 
\end{proposition} 
This preliminary result is an important step towards those which follow. Moreover, the result is of interest in and of itself since it applies provided that the number of factors is not underspecified, and irrespective of the relationship between $n$ and $T$, as long as both diverge to infinity. In contrast later in the paper, it will be required that the true number of factors is known, and that $n$ and $T$ grow in proportion (see Assumption \ref{ASS7}). Despite both the factors and the loadings having been concentrated out, the spaces spanned by both are implicitly estimated by their respective first order conditions, and as a result both $n$ and $T$ are required to diverge. The rate $a_{nT}$ is in line with the existing literature; see for example Theorem 4.1 in \cite{moon_linear_2015}, where a preliminary $\sqrt{\min\{n,T\}}$-consistency rate is established for a fixed number of (non-nuisance) parameters.\footnote{By imposing sparsity, and, with a judicious and data specific choice of penalty parameters, it may be possible to obtain faster rates of convergence. This may be of particular significance in very high dimensional settings with potentially $P > nT$, though such results are not pursued in this paper.}  

\subsection{Selection Consistency} \label{selcons}
In addition to the consistency result established in Proposition \ref{Prop2}, it is also desirable that the proposed estimator is selection consistent. This requires that, with probability approaching 1, the estimates of the truly zero coefficients are zero, while those of nonzero coefficients are nonzero.
\begin{Assumption} \label{ASS6} \color{white}.\color{black}\ 
$\min\{\gamma_{\rho},\gamma_{\beta}\} r_{nT}^{-\zeta} \rightarrow \infty$ as $n,T\rightarrow \infty$. \label{AS:8.1} 	
\end{Assumption} 
Assumption \ref{ASS6} ensures selection consistency of the estimator by taking advantage of the singularity of the penalty term at zero. Under Assumption \ref{ASS6},  $\min\{\gamma_{\rho},\gamma_{\beta}\} {|{\theta}^{\dagger}_p|^{-\zeta}}$ will be explosive in probability for those truly zero $\theta_p$ and as a result, asymptotically, the first order conditions cannot not be met unless $\hat{\theta}_p$ takes a value of exactly zero. For the following, recall from the end of Section \ref{sec model} that $\bs{\theta}_{(2)}$ contains the truly zero $\theta_p$.
\begin{proposition}[Selection Consistency]\label{Prop3}
Under Assumptions \ref{ASS1}--\ref{ASS6},
\begin{align}
\textup{Pr}\left( ||\bs{\hat{\theta}}_{(2)}||_2 = 0 \right) \rightarrow 1\ \text{as}\ n,T \rightarrow \infty.
\end{align}
\end{proposition}
Proposition \ref{Prop3} demonstrates that the estimator will correctly set coefficients with a true value of zero to exactly zero with probability approaching 1. Moreover, the consistency result proved in Proposition \ref{Prop2} implies that, with probability approaching 1, the estimates of nonzero coefficients must be nonzero. Thus together, Propositions \ref{Prop2} and \ref{Prop3} indicate that, with an appropriate choice of regularisation parameters, the PQMLE is model selection consistent. 

\subsection{Asymptotic Distribution}\label{asydis}
An implication of the model selection consistency result obtained in Proposition \ref{Prop3} is that the asymptotic distribution of the nonzero coefficient estimates coincides with that of the infeasible `oracle' estimator, which uses knowledge of which parameters are truly zero. The limiting distribution of the nonzero coefficient estimates is derived appealing to this result, and, in keeping with the high dimensional literature, this is done indirectly, by considering arbitrary linear combinations of parameters. As remarked in Section \ref{Penalty}, it is important for the validity of this approach that the true parameters have fixed values that are, by Assumption \ref{AS:4.0}, well separated from zero. If this were not the case then the finite sample distribution of the estimator could be quite different to that derived in Theorem \ref{Thrm1}; a point made clear by \cite{leeb}. However, this is a broader	issue in the literature and is particularly difficult to overcome in models of significant	complexity, where obtaining uniform results is often challenging.

\begin{Assumption}\label{ASS7} \color{white}-\color{black}\
\begin{enumerate}[label = 6.\arabic*] 
\item $\frac{P^5}{\min\{n,T\}} \rightarrow 0$ as $n,T \rightarrow \infty$. \label{AS:9.1}	
\item $\frac{1}{n} \bs{\Lambda}^{0'} \bs{\Lambda}^{0} \xrightarrow{p} \bs{\Sigma}_{\bs{\Lambda}^0}$ as $n \rightarrow \infty$ with $\bs{\Sigma}_{\bs{\Lambda}^0}$ being a $R^0 \times R^0$ positive definite matrix. \label{AS:9.2} 
\item $\frac{1}{T} \bs{F}^{0'} \bs{F}^{0} \xrightarrow{p} \bs{\Sigma}_{\bs{F}^0}$ as $T \rightarrow \infty$ with $\bs{\Sigma}_{\bs{F}^0}$ being a $R^0 \times R^0$ positive definite matrix. \label{AS:9.3}
\item $\frac{T}{n} \rightarrow c$ with $0 < c < \infty$. \label{AS:10.1}
\item $R = R^0$.  \label{AS:9.5}
\item $\max\{\gamma_{\rho},\gamma_{\beta}\} \sqrt{PnT} = o(1)$. \label{AS:10.2}
\end{enumerate}
\end{Assumption}
Assumption \ref{AS:9.1} ensures that the estimation of the
coefficients has a negligible effect on the estimation of the factors and
the loadings. \cite{lu_shrinkage_2016}, who consider estimation of a standard
regression model without interaction, require $P^2/\min\{n,T\}
\rightarrow 0$ for analogous purposes. A stronger condition is needed here
to ensure that the estimators of the reduced form factors $\bs{S}^{-1}(\bs{\rho}) \bs{\Lambda}$ converge sufficiently fast, since the reduced form is
implicitly used in instrumenting the endogenous variables. As $\bs{S}(%
\bs{\rho}) = \bs{I}_n - \sum^Q_{q=1} \rho_q \bs{W}_{\kern -0.2em q}$
involves an increasing number of weights matrices, the number of these
cannot be allowed to increase too quickly. Moreover the convergence of the covariance matrix requires further limits on the growth of $P$. 
\cite{peng_nonconcave_2004} require $P^5/n \rightarrow 0$, which corresponds to Assumption \ref{AS:9.1} in a cross-sectional framework. The condition given in 
\cite{liu_doctor_2017}, in a cross-sectional spatial model without a factor structure error
effects, also requires $P^5/n \rightarrow 0$. Assumptions %
\ref{AS:9.2} and \ref{AS:9.3} impose that the factors are strong, that is to
say that the factors and loadings have a nonnegligible impact on the
variance of the unobserved term $\bs{\eta} \coloneqq (\bs{\eta}_1,\ldots,\bs{\eta}_{T})$. Other authors consider models with weak factors however
this is not pursued here. Assumption \ref{AS:10.1} requires $n$ and $T$ to
grow in proportion. Similar asymptotic regimes are assumed in several papers in which biases arise in models with interactive fixed effects, and which use similar estimation approaches. Examples of these include \cite{moon_dynamic_nodate} and \cite{shi_spatial_2017}. Other papers, such as  \cite{bai_panel_2009} and \cite{lu_shrinkage_2016}, consider regimes where both $n/T^2, T/n^2 \rightarrow 0$, which provide similar limits on the relative growth rates of $n$ and $T$. Assumption \ref{AS:9.5} requires the true number of factors to be known. Nonetheless, Proposition \ref{Prop2} shows that the PQML
estimator remains consistent as long as the number of factors is not
understated; that is $R\geq R^0$. In the absence of interaction, \cite{moon_linear_2015} show that the asymptotic distribution of a least squares estimator is unaffected by overstatement of the number of factors, under certain conditions. It might, therefore, be expected that this extends to the present setting, however, since there may be significant complications in obtaining such results, the asymptotic distribution is derived under the assumption $R=R^0$. Section \ref{nofac} shows how the number of factors can be chosen consistently with reference to an information criterion. Assumption \ref{AS:10.2} strengthens the restrictions on the penalty term. 

Let $\mathcal{D}$ denote the sigma algebra generated by $\bs{\mathcal{X}}_1^*,\ldots,\bs{\mathcal{X}}_{K^*}^*$, $\bs{\Lambda}^0$ and $\bs{F}^0$. With a slight abuse of notation, in the following the subscripts $p$ and $q$ are also used to refer to an element in the indices $q = 1,\ldots,Q^0$ and $p = 1,\ldots,P^0$ which indexes quantities associated only with nonzero parameter values. Define $\bs{\bar{\mathscr{Z}}}_p \coloneqq \mathbb{E}[\bs{{\mathscr{Z}}}_p | \mathcal{D}]$, $\pmb{\mathbb{Z}}_p \coloneqq  \bs{M}_{\bs{\Lambda}^0}  \bs{\bar{\mathscr{Z}}}_p \bs{M}_{\bs{F}^0} + (\bs{\mathscr{Z}}_p - \bs{\bar{\mathscr{Z}}}_p)$, $\pmb{\mathbb{Z}}_{(1)} \coloneqq  (\text{vec}(\pmb{\mathbb{Z}}_1),\ldots,\text{vec}(\pmb{\mathbb{Z}}_{P^0}))$,
and $\bs{\mathcal{Z}}_{(1)} \coloneqq  (\text{vec}(\bs{\mathscr{Z}}_1),\ldots,\text{vec}(\bs{\mathscr{Z}}_{P^0}))$,
 that is, $\pmb{\mathbb{Z}}_{(1)}$ and $\bs{\mathcal{Z}}_{(1)}$ contain only covariates associated with nonzero parameters. Also, let
\begin{align}
\bs{D} &\coloneqq
\frac{1}{\sigma^2_0} \frac{1}{nT} 
\bs{\mathcal{Z}}_{(1)}'(\bs{M}_{\bs{F}^0} \otimes \bs{M}_{\bs{\Lambda}^0})\bs{\mathcal{Z}}_{(1)} 
+ 
\begin{pmatrix}
\bs{\Omega}
& 
\bs{0}_{Q^0 \times K^0}
\\
\bs{0}_{K^0 \times Q^0}
& 
\bs{0}_{K^0 \times K^0}
\end{pmatrix}, \label{Dm}
\end{align}
\begin{align}
\bs{V} 
&\coloneqq  \frac{\mathcal{M}^3_{\varepsilon}}{\sigma^4_0}
(\bs{\Phi} + \bs{\Phi}')
+
\frac{\mathcal{M}^{4}_{\vp} - 3 \sigma^4_0 }{\sigma^4_0}
\begin{pmatrix}
\bs{\Xi}  
& 
\bs{0}_{Q^0 \times K^0} \\
\bs{0}_{K^0 \times Q^0}
& 
\bs{0}_{K^0 \times K^0}
\end{pmatrix},
\label{SEC7.8}
\end{align}
where the matrices $\bs{\Omega}$ and $\bs{\Xi}$ are $Q^0 \times Q^0$ with elements $\Omega_{qq'} \coloneqq \frac{1}{n} \textup{tr}(\bs{G}_q (\bs{G}_{q'} + \bs{G}_{q'}')) - \frac{2}{n^2} \textup{tr}(\bs{G}_q) \textup{tr}(\bs{G}_{q'})$ and $\Xi_{qq'} \coloneqq \sum^T_{t=1} \sum^n_{i=1} (\bs{G}_q^*)_{ii} (\bs{G}_{q'}^*)_{ii}$, respectively, for $q,q' = 1,\ldots,Q^0$, and with $\bs{G}_q^* \coloneqq \bs{G}_q - \frac{1}{n} \textup{tr}(\bs{G}_q)\bs{I}_n$. The matrix $\bs{\Phi}$ is $P^0 \times P^0$ and has the structure $\bs{\Phi}\coloneqq(\bs{\bar{\Phi}}', \bs{0}_{P^0 \times K^0})'$, with $\bar{\Phi}_{qp} \coloneqq \sum^T_{t=1} \sum^n_{i=1}  (\pmb{\mathbb{Z}}_{p})_{it} (\bs{{G}}^*_{q})_{ii}$, for $q = 1,\ldots,Q^0$ and $p=1,\ldots,P^0$. 

\begin{Assumption}\label{ASS8} \color{white}-\color{black}\
\begin{enumerate}[label = 7.\arabic*] 
\item For some fixed integer $L$, $\pmb{\mathbb{S}}$ is a nonstochastic $L \times P^0$ matrix such that $\pmb{\mathbb{S}}\pmb{\mathbb{S}}'$ converges to a (entrywise) nonnegative matrix with eigenvalues bounded away from zero and infinity as $n,T \rightarrow \infty$. \label{AS:10.3}
\item There exist nonstochastic $P^0 \times P^0$ matrices $\pmb{\mathbb{D}} \coloneqq \mathbb{E}[\bs{D}]$ and $\pmb{\mathbb{V}} \coloneqq  \mathbb{E}[\bs{V}]$ such that $||\bs{D} - \pmb{\mathbb{D}}||_2= o_P(1), ||\bs{V} - \pmb{\mathbb{V}}||_2 = o_P(1)$, and the eigenvalues of $\pmb{\mathbb{D}}$, $\pmb{\mathbb{V}}$ and $\pmb{\mathbb{D}} + \pmb{\mathbb{V}}$ are bounded from below by zero and from above by a constant. \label{AS:10.4}
\end{enumerate}
\end{Assumption}
Since the limiting distribution of the estimator is difficult to derive directly, a
selection matrix $\pmb{\mathbb{S}}$ is introduced with a finite dimension $L$%
. Assumption \ref{AS:10.3} sets out basic properties of this matrix.
Assumption \ref{AS:10.4} ensures that the
covariance matrix of the PQMLE is well defined asymptotically. 
Let $\mathcal{M}^m_{\varepsilon}$ denote the $m$-th raw moment of ${\varepsilon}_{it}$, $\bs{J}_h \coloneqq (\bs{0}_{T \times (T-h)}, \bs{I}_T, \bs{0}_{T \times h})'$, are recall that $\bs{{\theta}}_{(1)}$ contains only those truly nonzero coefficients. 
\begin{theorem}[Asymptotic Normality]\label{Thrm1}
Under Assumptions \ref{ASS1}--\ref{ASS8},
\begin{align}
\sqrt{nT} \big(\pmb{\mathbb{S}}(\bs{D} + \bs{V})\pmb{\mathbb{S}}'\big)^{-\frac{1}{2}} \pmb{\mathbb{S}} \big( \bs{D} (\bs{\hat{\theta}}_{(1)} -  \bs{{\theta}}_{(1)}^0  ) - \pmb{\mathbbm{b}} \big) \xrightarrow[]{d} \mathcal{N}\big(\bs{0}_{L \times 1}, \bs{I}_{L}\big), \label{SEC7.6}
\end{align}
with
\begin{align}
\pmb{\mathbbm{b}} &\coloneqq 
\begin{pmatrix}
\pmb{\mathbbm{b}}^{(1)}\\
\bs{0}_{K^0 \times 1} \\
\end{pmatrix} 
+ 
\begin{pmatrix}
\pmb{\mathbbm{b}}^{(2)} \\
\bs{0}_{{K^*}^0\times 1} \\
\pmb{\mathbbm{b}}^{(3)}
\end{pmatrix}, \label{SEC7.7}
\end{align}
where the vector $\pmb{\mathbbm{b}}^{(1)}$ is $Q^0 \times 1$ with elements ${\mathbbm{b}}^{(1)}_q \coloneqq \sqrt{\frac{T}{n}}(\frac{R^0}{n}\textup{tr}(\bs{G}_q) - \textup{tr}(\bs{P}_{\bs{\Lambda}^0}\bs{G}_q) )  $, the vector $\pmb{\mathbbm{b}}^{(2)}$ is $Q^0 \times 1$ with elements ${\mathbbm{b}}^{(2)}_q \coloneqq  - \frac{1}{\sqrt{nT}} \sum^{T-1}_{h=1}\textup{tr}(\bs{J}_0 \bs{P}_{\bs{F}^0}\bs{J}_h')\textup{tr}(\bs{W}_{\kern -0.2em q} \bs{A}^{h}\bs{S}^{-1}) $ and the vector $\pmb{\mathbbm{b}}^{(3)}$ is $(Q^0 + 1) \times 1$ with first element ${\mathbbm{b}}^{(3)}_1 \coloneqq -\frac{1}{\sqrt{nT}} \sum^{T-1}_{h=1}\textup{tr}(\bs{J}_0 \bs{P}_{\bs{F}^0}\bs{J}_h')\textup{tr}( \bs{A}^{h-1}\bs{S}^{-1})$ and remaining elements  ${\mathbbm{b}}^{(3)}_{q+1} \coloneqq -\frac{1}{\sqrt{nT}} \sum^{T-1}_{h=1}\textup{tr}(\bs{J}_0 \bs{P}_{\bs{F}^0}\bs{J}_h')\textup{tr}(\bs{W}_{\kern -0.2em q} \bs{A}^{h-1}\bs{S}^{-1})$.\footnote{Note that here it is assumed that $\phi^0_{1}$ is nonzero so that ${\mathbbm{b}}^{(3)}_1$ appears in the bias term.} 
\end{theorem}
Theorem \ref{Thrm1} describes the asymptotic properties of the estimator for the nonzero
coefficients, detailing the asymptotic covariance matrix and the bias terms
which arise. Closer inspection reveals the bias $\pmb{\mathbbm{b}}^{(1)}$ is of order $\sqrt{T/n}$, while $\pmb{\mathbbm{b}}^{(2)}$ and $\pmb{\mathbbm{b}}^{(3)}$ are of order $\sqrt{n/T}$. These biases are a consequence of the incidental parameters in both dimensions of the panel. The bias $\pmb{\mathbbm{b}}^{(1)}$ is comprised to two parts. The first reflects the general loss of information in $\bs{G}_q$ resulting from reducing its rank by $R^0$ with the projection $\bs{M}_{\bs{\Lambda}^0}$. The second depends on the resemblance between the loadings and the network structure; both are sources of cross-sectional dependence and therefore may be conflated. If the column space of $\bs{G}_q$ is orthogonal
to the space of loadings, then $\bs{P}_{\bs{\Lambda}^0}\bs{G}_q = \bs{0}_{n
\times n}$ and the second part of $\pmb{\mathbbm{b}}^{(1)}$  does not feature. The second source of bias is characterised in $\pmb{\mathbbm{b}}^{(2)}$ for the $\bs{\rho}$ coefficients, and in $\pmb{\mathbbm{b}}^{(3)}$ for the $\bs{\phi}$ coefficients. These two biases arise due to the inclusion of a lagged outcome and are a generalisation of the usual fixed $T$ bias encountered in dynamic panels with individual fixed effects. As expected, when the number of parameters is fixed, with $\pmb{\mathbb{S}} = \bs{I}_{P^0}$ the distribution collapses to that of the QMLE where the covariance matrix has a typical sandwich form. 
\subsection{Bias Correction}

Given the characterisation of the bias term in Theorem \ref{Thrm1}, it is shown in the following proposition that this can be consistently
estimated and the limiting distribution of the PQMLE can be recentred. Let $\bs{\hat{D}}$ and $\pmb{\mathbbm{\hat{b}}}$ denote the analogues of $\bs{D}$ and $\pmb{\mathbbm{{b}}}$, respectively, where $\bs{\theta}^0, \bs{F}^0$, $\bs{\Lambda}^0$ and $\sigma^2_0$ are replaced by their estimates.
\begin{proposition}[Bias Correction]\label{Prop5}
Under Assumptions \ref{ASS1}--\ref{ASS8},
\begin{align}
\sqrt{nT} \big(\pmb{\mathbb{S}}(\bs{D} + \bs{V})\pmb{\mathbb{S}}'\big)^{-\frac{1}{2}}\pmb{\mathbb{S}}\bs{D}\big(\bs{\hat{\theta}}_{(1)}^c  - \bs{{\theta}}_{(1)}^0 \big) \xrightarrow[]{d} \mathcal{N}\big(\bs{0}_{L \times 1}, \bs{I}_{L}\big), \label{BIAS1}
\end{align}
with $\bs{\hat{\theta}}_{(1)}^c \coloneqq \bs{\hat{\theta}}_{(1)} - \bs{\hat{D}}^{-1}\pmb{\mathbbm{\hat{b}}}$ being the bias corrected estimator.  
\end{proposition}

\section{Implementation}\label{imp}
This section discusses a way in which the estimation procedure proposed in this paper can be implemented and, in particular, describes the approach used to obtain the results in Section \ref{ill}. This largely concerns how to choose the user specified inputs: the number of factors $R$, and the regularisation parameters  $\gamma_{\rho},\gamma_{\beta}$ and $\zeta$. Two methods to inform these choices are discussed in Sections \ref{penparam} and \ref{nofac}, with the overall suggestion being to proceed in the following way. First, by Proposition \ref{Prop2} the coefficients can be consistently estimated with knowledge only of an upper bound on the number of factors. Thus, with a suitable choice of the penalty parameters (discussed in Section \ref{penparam}) penalised estimation can be performed using a large $R$, and consistent estimates of the coefficients obtained. Using these coefficient estimates, a pure factor model can be constructed and the true number of factors detected (discussed in Section \ref{nofac}). Finally, the model should be re-estimated inputting the detected number of factors to obtain the final estimates. Of course, this multi-step procedure neglects to account for uncertainty at each stage and ideally it would be preferable to select both the penalty parameters and the number of factors jointly, however, the approach adopted here is pragmatic. Additional Monte Carlo results are provided in Appendix J of the Supplementary Material in order to assess the possible impact of varying the number of factors on the properties of the estimator.

\subsection{Choosing the Penalty Parameters}   
\label{penparam}
The fixed regularisation parameter ${\zeta}$ can typically be chosen in line with the rate of convergence of the initial estimator, in order to scale the parameter-specific weights $\omega_p$ appropriately. For example, if $r_{nT}$ is known to converge to zero slowly, $\zeta$ can be increased in order to ensure Assumptions \ref{ASS6} is satisfied.\footnote{In both the simulations and the application $\zeta$ is set equal to $4$, which performs well in practice and, with $\bs{\tilde{\theta}}$ as an initial estimate, would also be suitable for a general choice of $\gamma_{\rho}$ and $\gamma_{\beta}$; see footnote \ref{foot11}.} The other regularisation parameters $\gamma_{\rho}$ and $\gamma_{\beta}$, which must convergence to zero, could also be chosen simply as sequences which, in combination with $\zeta$, ensure Assumptions \ref{AS:4.1}, \ref{ASS6} and \ref{AS:10.2} are satisfied.\footnote{For example, if $r_{nT} = a_{nT}$, then $\zeta = 4$ and $\gamma_{\rho} = \gamma_{\beta} = 1/\min\{n,T\}$ would satisfy Assumptions \ref{AS:4.1} and \ref{ASS6} as long as $P^2/\min\{n,T\} \rightarrow 0$. With $n \propto T$ under Assumption \ref{AS:10.1}, and again, with $r_{nT} = a_{nT}$, then $\zeta = 4$ and $\gamma_{\rho} = \gamma_{\beta} = n^{-3/2}$ would satisfy Assumptions \ref{AS:4.1}, \ref{ASS6} and \ref{AS:10.2} as long as $P^4/\min\{n,T\} \rightarrow 0$.\label{foot11}} However, as an alternative, this section considers an information criterion that can be used to select $\gamma_{\rho}$ and $\gamma_{\beta}$, similar to what is proposed in \cite{lu_shrinkage_2016}. This is suggested in order to go some way in tailoring the choice of $\gamma_{\rho}$ and $\gamma_{\beta}$ to the data. Recalling $\bs{\gamma} \coloneqq (\gamma_{\rho},\gamma_{\beta})'$, the information criterion takes the form
\begin{align}
\textup{IC}^*(\bs{\gamma}) \coloneqq \hat{\sigma}^2(\bs{\gamma}) + \varrho_{\rho} | \mathcal{S}%
_{\rho}(\bs{\gamma}) | + \varrho_{\beta} | \mathcal{S}_{\beta}(\bs{\gamma})
|,  \label{IC}
\end{align}
where the notation $\hat{\sigma}^2(\bs{\gamma})$ is used for $ \hat{\sigma}^2(\bs{\hat{\theta}},\bs{\hat{\Lambda}})$ to emphasise the dependence on $\bs{\gamma}$, $\varrho_{\rho}$ and $\varrho_{\beta}$ are some positive penalty functions of $(n,T)$, $\mathcal{S}_{\rho}(\bs{\gamma})$, $\mathcal{S}_{\beta}(\bs{\gamma})$ denote the index sets for the nonzero elements of the parameter estimates under $\bs{\gamma}$, and $|\cdot|$ denotes the cardinality of a set. Following closely the exposition in \cite{lu_shrinkage_2016}, define $\mathcal{S}%
_{F,\rho} \coloneqq \{1,\ldots,Q\}$ and $\mathcal{S}_{F,\beta} \coloneqq \{1,\ldots,K\}$ as the index sets for the full set of weights matrices and for all covariates respectively. Analogous sets $\mathcal{S}_{T,\rho} \coloneqq \{1,\ldots,Q^0\}$ and $\mathcal{S}_{T,\beta} \coloneqq \{1,\ldots,K^0\}$ contain the indices of the relevant covariates and weights matrices. Next, define two closed intervals, $\Gamma_\rho \coloneqq [0,\bar{\gamma}_{\rho}]$ and $\Gamma_\beta \coloneqq [0, \bar{\gamma}_{\beta}]$, with $\Gamma_{\rho}, \Gamma_{\beta} \subset \mathbb{R}_+$ and where $\bar{\gamma}_{\rho}, \bar{\gamma}_{\beta}$ are two upper bounds beyond which all parameters would be set to zero. The space $\Gamma \coloneqq \Gamma_\rho \times \Gamma_\beta$ can be subdivided into three regions: 
\begin{align}
\Gamma^0 \coloneqq \{ \bs{\gamma} \in \Gamma : \mathcal{S}_{\rho}(\bs{\gamma}) = 
\mathcal{S}_{T,\rho}\ \text{and}\ \mathcal{S}_{\beta}(\bs{\gamma}) = 
\mathcal{S}_{T,{\beta}} \},  \nonumber
\end{align}
\begin{align}
\Gamma^{-} \coloneqq \{ \bs{\gamma} \in \Gamma : \mathcal{S}_{\rho}(\bs{\gamma})
\not\supseteq \mathcal{S}_{T,\rho}\ \text{or}\ \mathcal{S}_{\beta}(\bs{\gamma}%
) \not\supseteq \mathcal{S}_{T,\beta}) \},  \nonumber
\end{align}
\begin{align}
\Gamma^+ \coloneqq \{ \bs{\gamma} \in \Gamma : \mathcal{S}_{\rho}(\bs{\gamma})
\supset \mathcal{S}_{T,\rho}, \mathcal{S}_{\beta}(\bs{\gamma}) \supset 
\mathcal{S}_{T,\beta}\ \text{and}\ | \mathcal{S}_{\rho}(\bs{\gamma})| + | 
\mathcal{S}_{\beta}(\bs{\gamma})| > |\mathcal{S}_{T,\rho}| + |\mathcal{S}%
_{T,\beta}| \}.  \nonumber
\end{align}
Respectively, these are the
sets of $\bs{\gamma}$ in which the true model is selected, the model is
underfitted and the model is overfitted. The following assumptions are made. 
\begin{Assumption}\label{ASS9} \color{white}-\color{black}\
	\begin{enumerate}[label = 8.\arabic*] 
		\item $\frac{P^2}{\min\{n,T\}} \rightarrow 0$ as $n,T \rightarrow \infty$. \label{AS:12.0}	
		\item As $n,T \rightarrow \infty$, $(\sqrt{Q}a_{nT})^{-1} \varrho_{\rho}\rightarrow \infty, (\sqrt{Q}a_{nT})^{-1} \varrho_{\beta} \rightarrow \infty$, $Q^0 \varrho_{\rho}\rightarrow 0$, and ${K}^0 \varrho_{\beta} \rightarrow 0$.  \label{AS:12.1}
		\item For any $\bs{\gamma} \in {\Gamma}^{-} $, there exists $\sigma^2_{-}$ such that $\hat{\sigma}^2(\bs{\gamma}) \xrightarrow{p} \sigma^2_{-} > \sigma^2_0$.  \label{AS:12.2}
	\end{enumerate}
\end{Assumption}
Assumption \ref{ASS9} is analogous to Assumptions A.7 and A.8 in \cite{lu_shrinkage_2016}. Assumption \ref{AS:12.0} is required to ensure that for those $\bs{\gamma}$ which yield either the true model or an overfitted model, $\hat{\sigma}^2(\bs{\gamma})$ is consistent for $\sigma^2_0$. Assumption \ref{AS:12.1} requires that the penalty functions $\varrho_{\rho}$ and $\varrho_{\beta}$ relax sufficiently fast as sample size increases. In practice, there may be many functions which satisfy Assumption \ref{AS:12.1}, though these may have different impacts in finite samples; for further discussion see \cite{baiandng}. Assumption \ref{AS:12.2} ensures that underfitted models yield a larger mean squared error than a correctly fitted model. 
\begin{proposition}[Information Criterion Consistency]\label{Prop6} Under Assumptions \ref{ASS1}--\ref{ASS6} and \ref{ASS9},
	\begin{align}
	\textup{Pr}\left(\inf_{\bs{\gamma} \in \Gamma^- \cup \Gamma^+}  \textup{IC}^*(\bs{\gamma})  > \textup{IC}^*(\bs{\gamma}^0)   \right) \rightarrow 1\ \text{as}\ n,T \rightarrow \infty,
	\end{align} 
	for any $\bs{\gamma}^0 \in {\Gamma}^0$. 	
\end{proposition}

\subsection{Choosing the Number of Factors} \label{nofac} 
Following the procedure outlined at the beginning of Section \ref{imp}, penalised estimation can first be performed with the number of factors $R$ set to a large enough value, denoted by $R_{\max}$, in order to obtain consistent estimates of the parameters, denoted by $\check{\rho}$ and $\check{\beta}$. A pure factor model can then be constructed as
\begin{align}
	\bs{S}(\bs{\check{\rho}}) \bs{{Y}} - \sum^{K}_{k=1} \check{\beta}_k \bs{\mathscr{Z}}_k = \bs{\Lambda}^0 \bs{F}^{0^{\prime }} + \bs{\check{\varepsilon}}, \label{ICC}
\end{align}
with $\bs{\check{\varepsilon}} \coloneqq \sum^Q_{q=1}(\rho^0_q - \check{\rho}_q)\bs{G_\textit{q}} \big( \sum^{K}_{k=1} \beta^0_k \bs{\mathcal{X}}_k + \bs{\Lambda}^0 %
\bs{F}^{0^{\prime }} + \bs{\varepsilon} \big) + \sum^K_{k=1} (\beta^0_k -
\check{\beta}_k)\bs{\mathcal{X}}_k + \bs{\varepsilon}$. Existing information criteria can then be used to detect the number of factors, and this suggested number can be input into a second estimation step. For example, \cite{shi_spatial_2017} consider information criteria of the form 
\begin{align}
\textup{IC}(R) \coloneqq \log\left( \frac{1}{nT} \sum^{n}_{i = R + 1} \mu_i \left( \left( \bs{\Lambda}^0 \bs{F}^{0^{\prime }} + \bs{\check{\varepsilon}} \right) 
\left(\bs{\Lambda}^0 \bs{F}^{0^{\prime }} + \bs{\check{\varepsilon}}\right)' \right) \right) +
\varrho_{f} R, \label{aabb}
\end{align}
with $\varrho_{f}$ being a positive penalty function of $(n,T)$. With minor modification to Theorem 5 in that paper, it can be shown
that the information criterion in \eqref{aabb} is consistent in determining the number of factors, in the sense that $\lim_{n,T \rightarrow \infty} \textup{Pr} ( R^* = R^0 ) = 1$, with $R^* \coloneqq \argmin_{0 \leq R \leq R_{\max}}\textup{IC}(R)$ and under the additional assumption that the penalty function $\varrho_{f}$ satisfies $\varrho_{f} \rightarrow 0 $ and $a_{nT} \varrho_{f}
\rightarrow \infty$, with $a_{nT}$ being the preliminary rate established in Proposition \ref{Prop2}.

\section{Illustration}\label{ill}
This section demonstrates the finite sample performance and practicability of the procedure through the use of a small Monte Carlo study and an empirical example. 

\subsection{Simulations}\label{sim}
In the following design, the data are generated according to model \eqref{A1}, with the number of
parameters and weights matrices increasing with sample size. The design is summarised in Table \ref{tab1} with a little under half of the parameters taking a true value of $0$ for each sample size. Dashes in the table indicate that a covariate is absent. 
\begin{table}[H]
	\caption{True parameter values}
	\label{tab1}
	\begin{adjustbox}{width=15cm,center} 
		\begin{tabular}{cc|cccccccccccccccccccc}
			\cline{1-22}
			$n$ & $T$ & $\rho_1^0$ & $\rho_2^0$ & $\rho_3^0$ & $\rho_4^0$ & $\rho_5^0$ & $\delta_1^0$ & $\delta_2^0$ & $\delta_3^0$ & $\delta_4^0$ & $\delta_5^0$ & $\delta_{11}^0$ & $\delta_{12}^0$ & $\delta_{13}^0$ & $\delta_{14}^0$ & $\delta_{15}^0$ & $\phi_{1}^0$ & $\phi_{2}^0$ & $\phi_{3}^0$ & $\phi_{4}^0$ & $\phi_{5}^0$ \\
			\cline{1-22}
			& $25$  & $0.2$ & $0.2$ & $0$ & - & - & $3$ & $0$ & $-3$ & -   & -    & $1$ & $0$ & $-1$ & - & - & $0.15$ & $0$ & $-0.15$ & - & - \\
			$25$ & $50$  & $0.2$ & $0.2$ & $0$ & - & - & $3$ & $0$ & $-3$ & $0$ & -    & $1$ & $0$ & $-1$ & - & - & $0.15$ & $0$ & $-0.15$ & - & - \\
			& $100$ & $0.2$ & $0.2$ & $0$ & - & - & $3$ & $0$ & $-3$ & $0$ & $3$ & $1$ & $0$ & $-1$ & - & - & $0.15$ & $0$ & $-0.15$ & - & - \\
			\cline{1-22}
			& $25$  & $0.2$ & $0.2$ & $0$ & $0.2$ & - & $3$ & $0$ & $-3$ & -   & -    & $1$ & $0$ & $-1$ & $0$ & - & $0.15$ & $0$ & $-0.15$ & $0$ & - \\
			$50$ & $50$  & $0.2$ & $0.2$ & $0$ & $0.2$ & - & $3$ & $0$ & $-3$ & $0$ & -    & $1$ & $0$ & $-1$ & $0$ & - & $0.15$ & $0$ & $-0.15$ & $0$ & - \\
			& $100$ & $0.2$ & $0.2$ & $0$ & $0.2$ & - & $3$ & $0$ & $-3$ & $0$ & $3$ & $1$ & $0$ & $-1$ & $0$ & - & $0.15$ & $0$ & $-0.15$ & $0$ & - \\     
			\cline{1-22}
			& $25$  & $0.2$ & $0.2$ & $0$ & $0.2$ & $0$ & $3$ & $0$ & $-3$ & -   & -    & $1$ & $0$ & $-1$ & $0$ & $1$ & $0.15$ & $0$ & $-0.15$ & $0$ & $0$ \\
			$100$ & $50$  & $0.2$ & $0.2$ & $0$ & $0.2$ & $0$ & $3$ & $0$ & $-3$ & $0$ & -    & $1$ & $0$ & $-1$ & $0$ & $1$ & $0.15$ & $0$ & $-0.15$ & $0$ & $0$ \\
			& $100$ & $0.2$ & $0.2$ & $0$ & $0.2$ & $0$ & $3$ & $0$ & $-3$ & $0$ & $3$ & $1$ & $0$ & $-1$ & $0$ & $1$ & $0.15$ & $0$ & $-0.15$ & $0$ & $0$ \\    
			\cline{1-22}
		\end{tabular}%
	\end{adjustbox}
\end{table}
The error term $\varepsilon_{it}$, the loadings $\lambda_{ir}^0$ and the factors $%
f_{tr}^0$ are generated as standard normal variables.\footnote{For simplicity results are reported here only for idiosyncratic errors that are normally distributed. Similar results can be obtained under alternative error distributions and additional simulation results are available in Appendix J in the Supplementary Material.} Primitive exogenous variables are generated according to 
$x_{\kappa it}^* = \nu + \sum^{R^0}_{r=1} \lambda_{ir}^0 f_{rt}^0 + e_{it}$ with $\nu$ being uniformly drawn
from the integers $\{-10,\ldots,10\}$ and $e_{it} \sim \mathcal{N}(0,2)$. By design these are correlated with the factors and
the loadings and have associated coefficients $\delta^0_{\kappa}$. There are also additional covariates formed by interacting the $q$-th
weights matrix with the first primitive exogenous regressor in the
manner of $\eqref{A2}$. These covariates have associated coefficients $\delta_{1q}^0$. The number of weights matrices is increasing with $n$, with the first weights matrix
being constructed as if the cross-sectional units were
arrayed on a line and connected only to the units immediately to the left
and right. This is the simplest example of a path and produces a matrix with
ones along the diagonals directly above and below the main diagonal, and
zeros elsewhere. The remaining matrices are specified in similar fashion,
but now represent neighbours to the $q$-th degree. All matrices are then row normalised. Finally, a lag of outcomes is included, as well as interactions of this lagged outcome and the weights matrices.\footnote{Assumptions \ref{ASS1}--\ref{ASS9} are verified for this design in Appendix I of the Supplementary Material.}

Table \ref{tab2} reports bias corrected estimates $\bs{\hat{\theta}}^c$, across
various $n$ and $T$, each with $1000$ Monte Carlo replications, and where $R = R^0 = 3$.

\begin{table}[H]
	\caption{Bias of bias corrected estimates of nonzero parameters $(R=R^0)$}
	\label{tab2}
	\begin{adjustbox}{width=15cm,center} 
		\begin{tabular}{cc|rrrrrrrrrrr}
\cline{1-13}
$n$ & $T$ & \multicolumn{1}{c}{$\rho_1$} & \multicolumn{1}{c}{$\rho_2$} & \multicolumn{1}{c}{$\rho_4$} & \multicolumn{1}{c}{$\delta_1$} & \multicolumn{1}{c}{$\delta_3$} & \multicolumn{1}{c}{$\delta_5$} & \multicolumn{1}{c}{$\delta_{11}$} & \multicolumn{1}{c}{$\delta_{13}$} & \multicolumn{1}{c}{$\delta_{15}$} & \multicolumn{1}{c}{$\phi_{1}$} & \multicolumn{1}{c}{$\phi_{3}$} \\
\cline{1-13}
     & $25$  & $0.0002$ & $-0.0004$ &  - & $ 0.0008$ & $-0.0014$  & -        & $-0.0027$  & $0.0031$  & - & $-0.0004$  & $0.0004$  \\
$25$ & $50$  & $0.0001$ & $-0.0002$ &  - & $-0.0002$ & $-0.0006$  & -        & $-0.0016$  & $0.0026$  & - & $-0.0002$  & $0.0002$  \\
     & $100$ & $0.0001$ & $-0.0002$ &  - & $ 0.0001$ & $-0.0005$  & $0.0005$ & $-0.0014$  & $0.0017$  & - & $-0.0001$  & $0.0001$  \\
\cline{1-13}
     & $25$  & $0.0001$ & $0$       & $-0.0001$ & $ 0.0005$ & $ 0.0005$ & -        & $-0.0007$ & $ 0.0005$ & - & $-0.0002$ & $0.0002$  \\
$50$ & $50$  & $0.0002$ & $-0.0003$ & $0$       & $ 0.0002$ & $-0.0006$ & -        & $-0.0005$ & $ 0.0013$ & - & $-0.0001$ & $0.0001$  \\
     & $100$ & $0$      & $-0.0001$ & $0$       & $-0.0001$ & $-0.0003$ & $0.0003$ & $ 0$      & $ 0.0005$ & - & $-0.0002$ & $0.0002$  \\     
\cline{1-13}
     & $25$  & $-0.0001$ & $-0.0002$ & $0.0002$ & $0.0003$ & $-0.0011$ & -        & $-0.0004$ & $0.0022$ & $-0.0006$ & $-0.0003$ & $0.0003$ \\
$100$ & $50$  & $0$       & $0$       & $0$      & $0.0003$ & $-0.0002$ & -        & $ 0.0001$ & $0.0005$ & $-0.0006$ & $-0.0003$ & $0.0002$  \\
     & $100$ & $0.0001$  & $-0.0001$ & $0$      & $0$      & $-0.0001$ & $0.0002$ & $-0.0004$ & $0.0007$ & $-0.0001$ & $-0.0002$ & $0.0002$ \\    
	\cline{1-13}
		\end{tabular}%
	\end{adjustbox}
\end{table}

Table \ref{tab2} shows that the biases are generally decreasing with both $n$
and $T$ and tend to be larger for the parameters $\delta_1,\delta_3$ and $%
\delta_5$, as well as the exogenous spillovers $\delta_{11}, \delta_{13}$ and $\delta_{15}$. This is unsurprising since the covariates $\bs{\mathcal{X}}%
^*_{\kappa}$ are directly correlated with the loadings and the factors by
design. The biases of the $\rho_q$ parameters are lower since these
implicitly use the instrument $\bs{G}_q \bs{X}_t\bs{\beta}^0$, which may not itself be strongly correlated with the factors
and the loadings. The same is true of the coefficients $\phi_1$ and $\phi_3$, since the lags $\bs{\mathcal{Y}}_{-1}$ and interactions $\bs{W}_{\kern -0.2em q}\bs{\mathcal{Y}}_{-1}$ are less directly correlated with the factors and the loadings. These biases can be favourably compared with Table 6 in Appendix J in the Supplementary Material, which presents biases of the PQMLE without controlling for interactive effects, where there are large biases which persist with $n$ and $T$. 

\begin{table}[h]
	\caption{Coverage of nonzero parameter estimates $(R=R^0)$}
	\label{tab3}
	\begin{adjustbox}{width=14cm,center} 
		\begin{tabular}{cc|ccccccccccc}
	\cline{1-12}
	$n$ & $T$ & $\rho_1$ & $\rho_2$ & $\rho_4$ & $\delta_1$ & $\delta_3$ & $\delta_5$ & $\delta_{11}$ & $\delta_{13}$ & $\delta_{15}$ & $\phi_{1}$ & $\phi_{3}$ \\
	\cline{1-13}
	     & $25$  & $0.901$ & $0.902$ & - & $0.885$ & $0.908$ &  -      & $0.891$ & $0.897$ & - & $0.904$ & $0.907$  \\
	$25$ & $50$  & $0.906$ & $0.922$ & - & $0.921$ & $0.924$ &  -      & $0.922$ & $0.928$ & - & $0.916$ & $0.922$  \\
         & $100$ & $0.930$ & $0.919$ & - & $0.926$ & $0.929$ & $0.920$ & $0.917$ & $0.930$ & - & $0.929$ & $0.915$ \\
	\cline{1-13}
	     & $25$  & $0.920$ & $0.932$ & $0.931$ & $0.924$ & $0.927$ &    -    & $0.927$ & $0.927$ & - & $0.913$ & $0.920$  \\
    $50$ & $50$  & $0.939$ & $0.935$ & $0.931$ & $0.936$ & $0.926$ &    -    & $0.932$ & $0.917$ & - & $0.926$ & $0.930$  \\
	     & $100$ & $0.946$ & $0.942$ & $0.922$ & $0.932$ & $0.934$ & $0.932$ & $0.945$ & $0.921$ & - & $0.921$ & $0.928$  \\     
	\cline{1-13}
	     & $25$  & $0.929$ & $0.929$ & $0.923$ & $0.930$ & $0.921$ &    -    & $0.926$ & $0.916$ & $0.932$ & $0.934$ & $0.931$ \\
	$100$ & $50$  & $0.937$ & $0.935$ & $0.947$ & $0.941$ & $0.926$ &    -    & $0.920$ & $0.939$ & $0.939$ & $0.931$ & $0.934$  \\
     	 & $100$ & $0.947$ & $0.930$ & $0.942$ & $0.950$ & $0.946$ & $0.948$ & $0.941$ & $0.957$  & $0.942$ & $0.922$  & $0.921$ \\    
	\cline{1-13}
\end{tabular}%
	\end{adjustbox}
\end{table}

Table \ref{tab3} presents coverage probabilities of Wald confidence intervals based on Theorem \ref{Thrm1} and with a nominal coverage of $95\%$. These generally improve with 
$n$ and $T$, though due to the complexity of the design it is unsurprising that they do not do so monotonically.  Table $\ref{tab4}$ shows the percentage of true zero parameters correctly estimated as such, with the procedure performing well and achieving near 100\% accuracy across all $n$ and $T$. 
\begin{table}[H]
	\caption{Percentage of true zeros $(R=R^0)$}
	\label{tab4}
	\begin{adjustbox}{width=10cm,center} 
		\begin{tabular}{ccccccccccc}
\cline{1-11}
$n$ & $T$ & $\rho_3$ & $\rho_5$ & $\delta_2$ & $\delta_4$ & $\delta_{12}$ & $\delta_{14}$ & $\phi_{2}$ & $\phi_{4}$ & $\phi_{5}$ \\
\cline{1-11}
   & 25  & 99.9 & - & 100 &  -  & 99.9 & - & 99.9 & - & - \\
25 & 50  & 99.8 & - & 100 & 100 & 100  & - & 99.8 & - & - \\
   & 100 & 99.8 & - & 100 & 100 & 100  & - & 99.9 & - & - \\
\cline{1-11}
   & 25  & 100  & - & 100 &  -  & 100 & 100 & 100  & 100  & - \\
50 & 50  & 99.9 & - & 100 & 100 & 100 & 100 & 99.9 & 99.9 & - \\
   & 100 & 99.6 & - & 100 & 100 & 100 & 100 & 99.6 & 99.6 & - \\
\cline{1-11}
    & 25  & 99.9 & 99.9 & 99.9 &  -  & 99.9 & 99.9 & 99.9 & 99.9 & 99.9 \\
100 & 50  & 99.8 & 99.8 & 100  & 100 & 100  & 100  & 99.8 & 99.8 & 99.8 \\
    & 100 & 99.7 & 99.8 & 100  & 100 & 100  & 100  & 99.7 & 99.7 & 99.7 \\
\cline{1-11}
		\end{tabular}%
	\end{adjustbox}
\end{table}
The results reported in Tables \ref{tab2}--\ref{tab4} are computed with the correct number of factors inputted $(R = R^0 = 3)$, however, in practice, the true number of factors will not be known. To address this it was suggested in Section \ref{imp} to first perform penalised estimation of the model using an upper bound on the number of factors ($R = R_{\max}$) and then to construct a pure factor model and use the information criterion described in Section \ref{nofac} to detect the true number of factors. After this the model can be re-estimated inputting the detected number of factors to obtain the final estimates. In order to asses the effectiveness of this strategy, additional estimations are performed using an upper bound on the number of factors $R_{\max} = 6 > R^0$.\footnote{Table 15 in Appendix J provides additional results with $R_{\max} = 10$; the results are very similar.} A pure factor model is then constructed using these estimates and the information criterion \eqref{aabb} computed. Table \ref{tab45} presents the number of times, as a percentage, that the true number of factors is found to minimise the information criterion. Three variants of this criterion are used (IC1, IC2 and IC3) which differ only in their choice of penalty function $\varrho_{f}$.\footnote{The functions used in IC1, IC2 and IC3 are, respectively, $\log(\min\{n,T\})/\min\{n,T\}$, $((n+T)/(nT))\log(\min\{n,T\})$ and $((n+T)/(nT))\log((nT)/(n+T))$. For both $\varrho_{\rho}$ and $\varrho_{\beta}$ in IC$^*$, $\log(\min\{n,T\})/\min\{n,T\}$ is used.} As sample size increases, the performance of all three variants improves, though there is significant variability between the three criteria.\footnote{The penalty function IC1 is smaller in magnitude than IC2 and IC3 across all samples sizes. Moreover, unlike IC2 and IC3, IC1 only decreases when $\min\{n,T\}$ decreases. The overall result of this is under-penalisation for a larger $R$ and poor performance in smaller samples when $n = T$.} 
\begin{table}[H]
	\caption{True number of factors is selected \% ($R = R_{\max} = 6$)}
	\label{tab45}
	\begin{adjustbox}{width=10cm,center} 
		\begin{tabular}{c|ccc|ccc|ccc}
			\hline
			$T$   & \multicolumn{3}{c|}{25} & \multicolumn{3}{c|}{50} & \multicolumn{3}{c}{100} \\ \hline
			$n$   & IC1  & IC2  & IC3  & IC1  & IC2  & IC3  & IC1  & IC2  & IC3  \\ \hline
			25    & 0    & 96.5 & 79.2 & 46.1 & 99.6 & 99.4 & 99.9 & 99.9 & 99.9 \\
			50    & 43.8 & 99.1 & 98.8 & 7.3  & 100  & 100  & 100  & 100  & 100  \\
			100   & 99.7 & 99.8 & 99.8 & 100  & 100  & 100  & 99.9 & 100  & 100 \\ \hline
		\end{tabular}
	\end{adjustbox}
\end{table}
To gauge the likely impact of the factors not being known, estimation results with the number factors misspecified are provided in Appendix J in the Supplementary Material. These results illustrate cases in which the correct number of factors $R^0$ remains fixed at $3$, and yet $R = 1$, $R = 6$ and $R = 10$ are inputted in estimation. In line with the result in Proposition \ref{Prop2}, when the number of factors is underestimated ($R=1$) large biases persist, while the estimator remains consistent with the number of factors overestimated ($R=6$), even significantly so ($R=10$), though overestimation can result in considerable inefficiency.

\subsection{Application}
\label{app}
As an empirical demonstration, the method is applied to study the determinants of economic growth, using a panel data set where several countries are observed over multiple time periods. It is natural to suppose that economic growth might be influenced by unobserved shocks, as well as observable regressors, and in this spirit \cite{lu_shrinkage_2016} estimate a model of economic growth controlling for unobserved factors. In that paper, the authors focus, in particular, on applying shrinkage methods to determine an unknown number of factors. Extending their work to include interaction is well motivated, since one might reasonably expect the growth rates of different countries to be interrelated. Yet in such cases it can be difficult to specify weights matrices a priori. Indeed \cite{newgrow} remark: \textit{``Spatial methods may yet have an important role to play in growth econometrics. However, when these methods are adapted from the spatial statistics literature, they raise the problem of identifying the appropriate notion of space \ldots. countries are perhaps best thought of as occupying some general socio-economic-political space defined by a range of factors; spatial methods then require a means to identify their locations''}. The model studied in this paper may provide insight into growth rate determination, where uncertainty in specifying cross-national interactions provides an example of the type of uncertainty which the present methodology seeks to address. 

The data are obtained from \cite{lu_shrinkage_2016}, with additional data on income classifications from the World Bank. The outcome $y_{it}$ is the growth rate (Grth) in real GDP per capita for one of a cross-section of $108$ countries observed between the years 1970--2005. The same $9$ primitive exogenous covariates are used as in \cite{lu_shrinkage_2016}, which include variables such as life expectancy, population growth, and consumption, investment and government expenditure shares. A series of weights matrices are specified based on grouping countries according to four Word Bank classifications: high income ($\bs{W}_1$), upper-middle income ($\bs{W}_2$), lower-middle income ($\bs{W}_3$) and low income ($\bs{W}_4$) economies, and reflect the more general notion of a socio-economic space remarked upon on by \cite{newgrow}.  Each of these weights matrices are constructed by setting the $(i,j)$-th element to $1$ if country $i$ and $j$ share the same income classification, and setting it equal to zero otherwise, before then row normalising each of the matrices. 
\begin{table}[H]
	\caption{Estimation results without interaction.}
	\label{tab8}
	\begin{adjustbox}{width=15cm,center} 
		\begin{tabular}{c|ccccccccccc|ccc} 
			\cline{1-15}
			$R$    &        & Young & Fert  & Life  & Popu   & Invpri & Con    & Gov    & Inv    & Open   & Lag1  & IC1 & IC2 & IC3 \\
			\cline{1-15}	
			\multirow{2}{*}{$0$}  & estimate & $0$ & $0$ & $0$ & $-0.462$ & $0$  & $0$ & $0$ & $0.099$  & $0$  & $0.161$   & \multirow{2}{*}{$3.662$} & \multirow{2}{*}{$3.662$} & \multirow{2}{*}{$3.662$}  \\
			                      & t-stat   & $0$ & $0$ & $0$ & $-8.030$ & $0$  & $0$ & $0$ & $17.394$ & $0$  & $10.386$  &    &    &     \\
			\cline{1-15}	
			\multirow{2}{*}{$1$}  & estimate & $0$ & $0$ & $0$ & $-0.474$ & $0$ & $0$ & $-0.051$ & $0.118$  & $0$ & $0.137$ & \multirow{2}{*}{$3.508$} & \multirow{2}{*}{$3.541$} & \multirow{2}{*}{$3.531$} \\
			                      & t-stat   & $0$ & $0$ & $0$ & $-7.317$ & $0$ & $0$ & $-4.224$ & $18.504$ & $0$ & $8.855$ &    &    &    \\
			\cline{1-15}
			\multirow{2}{*}{$2$}  & estimate & $0$ & $0.444$ & $0$ & $-0.489$ & $0$ & $0$ & $-0.238$ & $0.228$  & $0$ & $0$ & \multirow{2}{*}{$3.449$} & \multirow{2}{*}{$3.515^{\dagger}$} & \multirow{2}{*}{$3.494$} \\ 
			                      & t-stat   & $0$ & $4.804$ & $0$ & $-5.186$ & $0$ & $0$ & $-9.424$ & $19.112$ & $0$ & $0$ &    &    &     \\
			\cline{1-15}	
			\multirow{2}{*}{$3$}  & estimate & $0$ & $0$ & $0$ & $-0.061$ & $0$ & $0$ & $-0.170$ & $0.228$  & $0$ & $0$  & \multirow{2}{*}{$3.420^{\dagger}$} & \multirow{2}{*}{$3.519$} & \multirow{2}{*}{$3.487^{\dagger}$} \\ 
		 		                  & t-stat   & $0$ & $0$ & $0$ & $-0.690$ & $0$ & $0$ & $-8.644$ & $19.821$ & $0$ & $0$  &    &    &     \\
			\cline{1-15}
			\multirow{2}{*}{$6$}  & estimate & $0$ & $0.165$ & $0$ & $-0.432$ & $0$ & $0$ & $-0.174$ & $0.217$  & $0$ & $0$ & \multirow{2}{*}{$3.437$} & \multirow{2}{*}{$3.636$} & \multirow{2}{*}{$3.572$} \\ 
			                      & t-stat   & $0$ & $2.131$ & $0$ & $-4.393$ & $0$ & $0$ & $-7.779$ & $19.524$ & $0$ & $0$ &    &    &     \\
			\cline{1-15}
			\cline{1-15}
		\end{tabular}%
	\end{adjustbox}
\end{table}	
Table \ref{tab8} reports bias corrected estimates $\bs{\hat{\theta}}^c$ in the absence of interaction.\footnote{Note that, in the absence of interaction, the quasi-maximum likelihood estimator reduces to the usual principal component least squares estimator \citep[e.g.,][]{bai_panel_2009}.} Three variants (IC1, IC2 and IC3) of the information criterion given in \eqref{aabb}  are computed using estimates generated inputting $R = R_{\max} = 6$.\footnote{These variants are the same as those used in simulations.} In two out of three cases, the information criteria suggest that the number of factors $R$ is $3$, matching the number suggested in \cite{lu_shrinkage_2016}. The estimates corresponding to $R = 3$ can be compared to the results for the AgLasso (which selects $R=3$) given in Table 7 of \cite{lu_shrinkage_2016}. In this case coefficient estimates and t-statistics are similar. 
\begin{table}[H]
	\renewcommand{\thetable}{\arabic{table}}
	\caption{Estimation results with endogenous interaction and temporal lags.}
	\label{tab10a}
	\begin{adjustbox}{width=15cm,center} 
		\begin{tabular}{c|cccccccccccccc}
			\cline{1-15}
			$R$  &  & $\bs{W}_{\kern -0.2em 1}\times$Grth & $\bs{W}_{\kern -0.2em 2}\times$Grth & $\bs{W}_{\kern -0.2em 3}\times$Grth & $\bs{W}_{\kern -0.2em 4}\times$Grth & Young & Fert & Life & Popu & Invpri & Con & Gov & Inv & Open \\
			\cline{1-15}	
			\multirow{2}{*}{$0$}  & estimate &  $0.210$ & $0.150$ & $0$ & $0.258$ & $0$ & $0$ & $0$ & $-0.492$ & $0$ & $0$ & $0$ & $0.090$  & $0$ \\
			     & t-stat   &  $3.167$ & $1.297$ & $0$ & $3.795$ & $0$ & $0$ & $0$ & $-8.460$ & $0$ & $0$ & $0$ & $15.115$ & $0$  \\
			\cline{1-15}
			\multirow{2}{*}{$1$}  & estimate & $0.295$ & $0.289$ & $-0.192$ & $0.345$ & $0$ & $-0.070$ & $0$ & $-0.443$ & $0$ & $0$ & $-0.050$ & $0.111$  & $0$ \\
			     & t-stat   & $3.688$ & $4.060$ & $-1.475$ & $5.141$ & $0$ & $-1.239$ & $0$ & $-4.977$ & $0$ & $0$ & $-4.511$ & $16.372$ & $0$  \\
			\cline{1-15}
			\multirow{2}{*}{$2$}  & estimate & $0.100$ & $0$ & $-0.325$ & $0.207$ & $0$ & $0.355$  & $0$ & $-0.477$ & $0$ & $0$ & $-0.237$ & $0.218$  & $0$ \\
			     & t-stat   & $1.323$ & $0$ & $-2.383$ & $2.808$ & $0$ & $3.958$  & $0$ & $-5.107$ & $0$ & $0$ & $-9.493$ & $17.823$ & $0$ \\			
			\cline{1-15}
			\multirow{2}{*}{$3$}  & estimate & $0.195$ & $0$ & $-0.305$ & $0.227$ & $0$ & $-0.001$ & $0$ & $-0.095$ & $0$ & $0$ & $-0.188$ & $0.215$  & $0$ \\
			     & t-stat   & $2.603$ & $0$ & $-2.277$ & $3.099$ & $0$ & $-0.016$ & $0$ & $-0.953$ & $0$ & $0$ & $-8.129$ & $18.055$ & $0$ \\
			\cline{1-15}
			\multirow{2}{*}{$6$}  & estimate & $0$ & $0$ & $-0.202$ & $0$ & $0.093$ & $-0.946$ & $0$ & $-0.570$ & $0$ & $0$ & $-0.225$ & $0.220$  & $0$ \\
			     & t-stat   & $0$ & $0$ & $-2.224$ & $0$ & $6.179$ & $-4.760$ & $0$ & $-5.703$ & $0$ & $0$ & $-8.536$ & $16.954$ & $0$ \\
			\cline{1-15}
		\end{tabular}%
	\end{adjustbox}
\end{table}
\begin{table}[H]
	  \addtocounter{table}{-1}
	\renewcommand{\thetable}{\arabic{table} Continued}
	\caption{Estimation results with endogenous interaction and temporal lags.}
	\label{tab10b}
	\begin{adjustbox}{width=10cm,center} 
		\begin{tabular}{c|cccccc|ccc}
			\cline{1-10}
			$R$  &   & Lag1 & $\bs{W}_{\kern -0.2em 1}\times$Lag1 & $\bs{W}_{\kern -0.2em 2}\times$Lag1 & $\bs{W}_{\kern -0.2em 3}\times$Lag1 & $\bs{W}_{\kern -0.2em 4}\times$Lag1 & IC1 & IC2 & IC3 \\
			\cline{1-10}	
			\multirow{2}{*}{$0$}    & estimate & $0.159$ & $0$ & $0$ & $0$ & $0$ & \multirow{2}{*}{$3.723$} & \multirow{2}{*}{$3.723$} & \multirow{2}{*}{$3.723$} \\
			                        & t-stat   & $10.279$  & $0$ & $0$ & $0$ & $0$ &  &  &  \\
			\cline{1-10}
			\multirow{2}{*}{$1$}    & estimate & $0.129$ & $0.172$ & $0$ & $0.400$ & $0$ & \multirow{2}{*}{$3.496$} & \multirow{2}{*}{$3.529$} & \multirow{2}{*}{$3.519$} \\
			                        & t-stat   & $8.145$ & $1.730$ & $0$ & $2.695$ & $0$ &  &  &  \\
			\cline{1-10}
			\multirow{2}{*}{$2$}    & estimate & $0.031$ & $0$ & $0$ & $0.177$ & $0$ & \multirow{2}{*}{$3.442$} & \multirow{2}{*}{$3.508^\dagger$} & \multirow{2}{*}{$3.487$} \\
			                        & t-stat   & $1.965$ & $0$ & $0$ & $1.137$ & $0$ &  &  &  \\		
			\cline{1-10}
			\multirow{2}{*}{$3$}    & estimate & $0.033$ & $0$ & $0$ & $0.233$ & $0$ & \multirow{2}{*}{$3.417^\dagger$} & \multirow{2}{*}{$3.516$} & \multirow{2}{*}{$3.484^\dagger$} \\
			                        & t-stat   & $2.070$ & $0$ & $0$ & $1.572$ & $0$ &  &  &  \\
			\cline{1-10}
			\multirow{2}{*}{$6$}    & estimate & $0$ & $0$ & $0$ & $0$ & $0$ & \multirow{2}{*}{$3.433$} & \multirow{2}{*}{$3.632$} & \multirow{2}{*}{$3.568$} \\
			                        & t-stat   & $0$ & $0$ & $0$ & $0$ & $0$ &  &  &  \\
			\cline{1-10}
		\end{tabular}%
	\end{adjustbox}
\end{table}
Table \ref{tab10a} reports estimation results once endogenous interaction and dynamic interaction is added. Government spending and investments shares in particular remain highly significant. However there is also evidence to suggest that there are significant endogenous spillovers, especially between high income and low income countries. The results indicate that amongst these two groups of countries, growth rates are interrelated with a positive spillover. In addition, there is evidence to suggest the presence of dynamic spillovers, these being positive, between lower-middle income countries.  

\section{Conclusion}\label{concfl}
To conclude, this paper considers the estimation of a model of cross-section interaction, whose salient features are a potentially increasing number of weights matrices and a factor structure in the error term. A penalised quasi-maximum likelihood estimator is proposed, in order to perform inference on network spillovers of various kinds, and its asymptotic properties are studied. A small Monte Carlo study reports good finite sample performance, and an empirical application studying the determinants of economic growth finds positive spillovers between the growth rates of high income and low income countries.     
 
This work could be extended in several directions. For instance, one might consider possible endogeneity of the weights matrices as in \cite{endo} and \cite{prucha}, or extend the use of weights matrices to the error term. Since they are observed, the possibility of time varying weights matrices might also be of interest. With some modified assumptions, the consistency result in Proposition \ref{Prop2} could be extended quite readily to this case, though additional work would be required to characterise the asymptotic distribution. Another prospect might be to consider higher dimensional settings, for example, one might consider an entirely unknown weights matrix, modelled in this framework as a series of weights matrices containing a single unitary element. However, identification in this setting would need to be carefully studied since including parameters which increase too quickly with $n$, alongside the factor loadings, may present complications. As a final thought, it might also be natural to allow the number of factors to increase with sample size. When the number of interacting cross-sectional units increases, and more units in a network are observed, it might be expected that additional latent structures in the error term would lead to an increase in the rank of the factor term. 
\\

%%%%%%%%%%    Appendies    %%%%%%%%%%%
\appendix\markboth{Appendix}{Appendix}
\renewcommand{\thesection}{\Alph{section}}
\numberwithin{equation}{section}
\begin{appendices}

\section{Proofs of Main Results}\label{mainres}	

This appendix provides proofs of the main results before which a series of lemmas are stated. The proofs of these lemmas are given in the Supplementary Material. The following facts are used repeatedly \citep[proofs can be found, for instance, in][]{moon_dynamic_nodate}. Let $\bs{A}$ and $\bs{B}$ be two conformable matrices. Then $||\bs{A}||_2 \leq ||\bs{A}||_F \leq \sqrt{\text{rank}(\bs{A})} ||\bs{A}||_2$, $||\bs{A}||_2 \leq \sqrt{||\bs{A}||_1 ||\bs{A}||_{\infty}}$ and $||\bs{A} \bs{B}||_F \leq ||\bs{A}||_F ||\bs{B}||_2 \leq ||\bs{A}||_F ||\bs{B}||_F$. Let the $i$-th row of an $n \times m$ matrix $\bs{B}$ be denoted $(\bs{B})_{i \bcdot }$, and the $j$-th column be denoted $(\bs{B})_{\bcdot  j}$. Then $\big( \sum^{m}_{j=1}||\bs{B}_{\bcdot j}||_2^2 \big)^\frac{1}{2} = \big( \sum^{n}_{i=1}||\bs{B}_{i \bcdot}||_2^2 \big)^\frac{1}{2} =  ||\bs{B}||_F$. Finally, under Assumption \ref{AS:1.1}, $||\bs{\varepsilon}||_2 = O_P(\sqrt{\min\{n,T\}})$ \citep[see][]{latala_estimates_nodate}. 

\textbf{Estimated factors and loadings}:
The maximiser of ${\mathcal{Q}}(\bs{\theta},\bs{\Lambda})$ with respect to $\bs{\Lambda}$ is not unique, since for any $\bs{\Lambda}^* = \bs{\Lambda}\bs{H}$, with $\bs{H}$ being an $R \times R$ invertible matrix, $\bs{M}_{\bs{\Lambda}} = \bs{M}_{\bs{\Lambda}^*}$. In order to achieve uniqueness of the estimators of $\bs{\Lambda}$ and $\bs{F}$, the normalisations that $\frac{1}{n}\bs{\Lambda}'\bs{\Lambda} = \bs{I}_{R}$ and $\bs{F}'\bs{F}$ is a diagonal matrix are adopted, see for example \cite{bai_panel_2009}.\footnote{%
	It is straightforward to see that such matrices exist. For example, by the singular value decomposition, decompose $\bs{\Lambda}\bs{F}' = \bs{U}\bs{S}\bs{V}'$. Let $\bs{\check{\Lambda}}$ be the $R$ columns of $\sqrt{n}\bs{U}$ associated with the nonzero singular values and $\bs{\check{F}}'$ be the corresponding $R$ rows of $\bs{S}\bs{V}'/\sqrt{n}$. As the columns of $\bs{U}$ and $\bs{V}$ are orthogonal, and $\bs{S}$ is diagonal, it follows that $\bs{\check{\Lambda}}'\bs{\check{\Lambda}}/n = \bs{I}_R$, $\bs{\check{F}}'\bs{\check{F}}$ is diagonal and $\bs{\check{\Lambda}}\bs{\check{F}}' = \bs{\Lambda}\bs{F}'$.} 
Under these normalisations, define
\begin{align}
\bs{\hat{\Lambda}}(\bs{\theta})
&\coloneqq \argmin_{\bs{\Lambda}: \frac{1}{n} \bs{\Lambda}'\bs{\Lambda} = \bs{I}_R } \left\{ \frac{1}{nT} \sum^T_{t=1} \bs{e}_t^{\prime}
\bs{M}_{\bs{\Lambda}} \bs{e}_t   \right\} 
= \argmax_{\bs{\Lambda}: \frac{1}{n} \bs{\Lambda}'\bs{\Lambda} = \bs{I}_R } \left\{  \frac{1}{n} \text{tr}  \left( \bs{\Lambda}' \frac{1}{nT}\sum^T_{t=1}   \bs{e}_t \bs{e}_t' \bs{\Lambda} \right) \right\}. \label{lhat}
\end{align}
It can be shown that the columns of $\bs{\hat{\Lambda}}(\bs{\theta})$ are equal to $R$ orthonormal eigenvectors of the matrix $\frac{1}{nT} \sum^T_{t=1} \bs{e}_t^{\prime} \bs{e}_t$ associated with the $R$ largest eigenvalues. With $\bs{\hat{F}}'\bs{\hat{F}}$ being diagonal, $\bs{\hat{\Lambda}}(\bs{\theta})$ will be unique, up to a permutation of its rows and a column-wise change of sign, provided the diagonal entries of $\bs{\hat{F}}'\bs{\hat{F}}$ are distinct. Hereafter $\bs{\hat{\Lambda}} \coloneqq \bs{\hat{\Lambda}}(\bs{\hat{\theta}})$. 

\textbf{Additional notation}:
For a matrix (and implicitly also for a vector) $\bs{B}$, $\bs{B} = \bs{O}_P (a_{nT})$ means that $||\bs{B}||_2 = O_P(a_{nT})$. Similarly $\bs{B} = \bs{o}_P (a_{nT})$ means that $||\bs{B}||_2 = o_P(a_{nT})$. The elements of the matrices $\bs{\mathcal{X}}_{\kappa}^*$, $\bs{\mathcal{X}}_k$, $\bs{\mathscr{Z}}_p$, $\bs{\varepsilon}$, $\bs{\Lambda}$ and $\bs{F}$ are respectively denoted $x_{\kappa it}^*$, $x_{kit}$, $z_{pit}$, $\varepsilon_{it}$, $\lambda_{ir}$ and $f_{tr}$. For any other $n \times m$ matrix $\bs{B}$, the $(i,j)$-th element is denoted $(\bs{B})_{ij}$. Finally, the $l$-th raw moment of some random variable $s$ is denoted  $\mathcal{M}^l_{s}$.

\renewcommand{\thesubsection}{\normalsize Lemma \thesection.\arabic{subsection}}
\makeatletter
\def\@hangfrom#1{\setbox\@tempboxa\hbox{{#1}}%
	\hangindent 0pt%\wd\@tempboxa
	\noindent\box\@tempboxa}
\makeatother
\subsection{\normalfont\normalsize \textit{For any $n \times n$ diagonalisable positive definite matrix $\bs{B}$, $\textup{det}(\bs{B})^{\frac{1}{n}} \leq  \frac{1}{n} \textup{tr}(\bs{B})$, with equality if and only if $\bs{B} = c\bs{I}_n$ for some $c > 0$.}}\label{firstlem}
\renewcommand\theenumi{\thesection.\arabic{subsection}}
\renewcommand\labelenumi{} 

\subsection{\normalfont\normalsize\textit{Under Assumptions} \ref{ASS1}--\ref{ASS2},}
\renewcommand\labelenumi{(\roman{enumi})}
\renewcommand\theenumi{\thesection.\arabic{subsection}(\roman{enumi})}
\begin{enumerate}\small
\item $\bs{S}(\bs{\rho}) \bs{S}^{-1} = \bs{I}_n + \sum^{Q}_{q=1}(\rho^0_q - \rho_q)\bs{G}_q$; \label{L1.2}
\item $||\bs{\mathscr{Z}}_p||_2 \leq ||\bs{\mathscr{Z}}_p||_F = O_P(\sqrt{nT})$ \textit{for} $p = 1,\ldots,P$; \label{L1.9}
\item $||\bs{\Lambda}^0||_2 \leq ||\bs{\Lambda}^0||_F = O_P(\sqrt{n}), ||\bs{F}^0||_2 \leq ||\bs{F}^0||_F = O_P(\sqrt{T})$; \label{L1.10}
\item $(\sum^{P}_{p=1}||\bs{\mathscr{Z}}_p||_2^2)^{\frac{1}{2}}, (\sum^{T}_{t=1}||\bs{Z}_t||_2^2)^{\frac{1}{2}}  = O_P(\sqrt{PnT})$; \label{L1.12}
\item $\mathbb{E}\big[\sum^{P}_{p=1}\text{tr}(\bs{\mathscr{Z}}_p'{\bs{S}(\bs{\rho})}\bs{S}^{-1}\bs{\varepsilon})^2\big]= O(PnT)$; \label{L1.15}
\item $||\bs{\varepsilon}||_F = O_P(\sqrt{nT})$; \label{L1.16}
\item $(\sum^T_{t=1} ||\bs{{X}}_t \bs{\beta}^0||_2^2 )^{\frac{1}{2}} = O_P(\sqrt{nT})$; \label{L1.17}
\item $||\bs{S}(\bs{{\rho}}) \bs{S}^{-1} - \bs{I}_n||_2 =   O_P(\sqrt{Q}||\bs{\theta}^0 - \bs{{\theta}}||_2)$. \label{L2.1}
\end{enumerate}

\subsection{\normalfont \normalsize \textit{Under Assumptions} \ref{ASS1}--\ref{ASS4},}
\renewcommand\labelenumi{(\roman{enumi})}
\renewcommand\theenumi{\thesection.\arabic{subsection}(\roman{enumi})}
\begin{enumerate}\small
\item $(\frac{1}{nT}\sum^T_{t=1}||\bs{Z}_t(\bs{\theta}^0 - \bs{{\theta}})||^2_2 )^{\frac{1}{2}} = O_P(||\bs{\theta}^0 - \bs{{\theta}}||_2)$; \label{L1.18}
\item $\hat{\sigma}^{-2}(\bs{\hat{\theta}}, \bs{{\Lambda}}) = O_P(1)$. \label{L1.19}
\end{enumerate}

\subsection{\normalfont\normalsize \textit{Under Assumptions} \ref{ASS1}--\ref{ASS7},}\label{Prop4}
\begin{align}
\bs{D}
\sqrt{nT} (\bs{\hat{\theta}}_{(1)} - \bs{\theta}^0_{(1)})
=&\ \frac{1}{\sigma^2_0} \frac{1}{\sqrt{nT} }\bs{\mathcal{Z}}_{{(1)}}' (\bs{M}_{\bs{F}^0} \otimes \bs{M}_{\bs{\Lambda}^0}) \textup{vec}(\bs{\varepsilon}) \notag \\
&+\ \frac{1}{\sigma^2_0} \frac{1}{\sqrt{nT}}
\begin{pmatrix}
\text{tr} \left( (\bs{G}_1^* \bs{\varepsilon} )' \bs{M}_{\bs{\Lambda}^0} \bs{\varepsilon} \bs{M}_{\bs{F}^0} \right) \\
\vdots \\
\text{tr} \left ( (\bs{G}_{Q^0}^*  \bs{\varepsilon} )' \bs{M}_{\bs{\Lambda}^0} \bs{\varepsilon} \bs{M}_{\bs{F}^0} \right) \\
\bs{0}_{K^0 \times 1} 
\end{pmatrix} 
+ \bs{o}_P(1), \notag 
\end{align}
\textit{where the matrix $\bs{D}$ is defined in equation \eqref{Dm} and the matrices $\bs{G}_q^*$, $q = 1,\ldots,Q^0$, are those associated with nonzero coefficients.}

\subsection{\normalfont \normalsize \textit{Under Assumptions} \ref{ASS1}--\ref{ASS7},}
\renewcommand\labelenumi{(\roman{enumi})}
\renewcommand\theenumi{\thesection.\arabic{subsection}(\roman{enumi})}
\begin{enumerate}\small

\item $||\bs{D}^{-1} - \bs{\hat{D}}^{-1}||_2 = O_P(({Q^0})^{1.5}P^0|| \bs{\theta}^0 - \bs{\hat{\theta}}||_2) + O_P\left( \frac{Q^0P^0}{\sqrt{\min\{n,T\}}} \right)$; \label{sub11}

\item $\mathbb{E}\left[ \sum^{Q^0}_{q=1} \left( \text{tr}((\bs{G}^*_q\bs{\varepsilon})' \bs{P}_{\bs{\Lambda}^0}\bs{\varepsilon}) - \sigma^2_0 T  \text{tr} (\bs{P}_{\bs{\Lambda}^0} \bs{G}^*_q) \right)^2 \right] = O(Q^0T)$;  \label{aaadd3}

\item $\mathbb{E}\left[ \sum^{Q^0}_{q=1} \left( \text{tr}((\bs{G}^*_q\bs{\varepsilon})' \bs{P}_{\bs{\Lambda}^0}\bs{\varepsilon}\bs{P}_{\bs{F}^0} ) - \sigma^2_0 R^0 \text{tr}(\bs{P}_{\bs{\Lambda}^0}\bs{G}^*_q) \right)^2 \right] = O(Q^0)$; \label{aaadd4}

\item $\mathbb{E}\left[ \sum^{Q^0}_{q=1} \left( \text{tr}((\bs{G}^*_q\bs{\varepsilon})' \bs{\varepsilon} \bs{P}_{\bs{F}^0}) - \sigma^2_0 R^0 \text{tr}(\bs{G}^*_q) \right)^2 \right] = O(Q^0n)$; \label{aaadd5}
\item \parbox[t]{\textwidth}{
	\vspace{-2.2em}
\begin{align}
\frac{1}{\sigma^2_0} \frac{1}{\sqrt{nT} }
\begin{pmatrix}
	\text{tr} \left( ( \bs{\mathscr{Z}}_1 - \bs{\bar{\mathscr{Z}}}_1 )' ( \bs{P}_{\bs{\Lambda}^0} \bs{\varepsilon} + \bs{M}_{\bs{\Lambda}^0}  \bs{\varepsilon} \bs{P}_{\bs{F}^0} ) \right) \\
	\vdots \\
	\text{tr} \left ( ( \bs{\mathscr{Z}}_{P^0} - \bs{\bar{\mathscr{Z}}}_{P^0} )' ( \bs{P}_{\bs{\Lambda}^0} \bs{\varepsilon} + \bs{M}_{\bs{\Lambda}^0}  \bs{\varepsilon} \bs{P}_{\bs{F}^0} )\right) 
\end{pmatrix}
=
\begin{pmatrix}
\pmb{\mathbbm{b}}^{(2)} \\
\bs{0}_{{K^*}^0 \times 1} \\
\pmb{\mathbbm{b}}^{(3)}
\end{pmatrix}+ \bs{o}_P(1), \notag 
\end{align}}
\label{aaadd6}
\end{enumerate}\vspace{-0.7cm}
\textit{where the matrices $\bs{G}_q^*$, $q = 1,\ldots,Q^0$, and the variables $\bs{\mathscr{Z}}_p - \bs{\bar{\mathscr{Z}}}_p$, $p = 1,\ldots,P^0$, are those associated with nonzero coefficients.}

\subsection{\normalfont\normalsize \textit{Under Assumptions} \ref{ASS1}--\ref{ASS8}, $\frac{1}{\sqrt{nT}} \frac{1}{\sigma^2_0} \left( \pmb{\mathbb{S}} (\pmb{\mathbb{D}}  + \pmb{\mathbb{V}}) \pmb{\mathbb{S}}' \right)^{-\frac{1}{2}} \pmb{\mathbb{S}}  \bs{c}
	\xrightarrow{d} \mathcal{N}(\bs{0}_{L \times 1}, \bs{I}_{L})$,\ \textit{where}\ $\bs{c} \coloneqq \pmb{\mathbb{Z}}_{(1)}' \textup{vec}(\bs{\varepsilon})
	+ 
	(\text{tr}(\bs{\varepsilon}'\bs{G}^*_{1}\bs{\varepsilon}),\ldots,
	\text{tr}(\bs{\varepsilon}'\bs{G}^*_{Q^0}\bs{\varepsilon}),\bs{0}_{1 \times K^0})'$, the matrices $\pmb{\mathbb{S}},\pmb{\mathbb{D}}$\ \textit{and}\ $\pmb{\mathbb{V}}$\ \textit{are defined in Assumptions \ref{AS:10.3} and \ref{AS:10.4}, and the matrices}\ $\bs{G}_q^*,$ $q = 1,\ldots,Q^0$,\ \textit{those associated with}\ \textit{nonzero coefficients.} }\label{aaadd2}  
\renewcommand\labelenumi{(\roman{enumi})}
\renewcommand\labelenumi{}

\begin{pff}[Proof of Proposition \ref{Prop2}] 
Here only a sketch of the proof is provided. A more detailed version can be found in Appendix D of the Supplementary Material. \\
\textbf{Consistency of the QMLE $\bs{\tilde{\theta}}$}\\
First, consider the average concentrated quasi-likelihood 
\begin{IEEEeqnarray}{rCl}
\mathcal{L}(\bs{\theta}) \coloneqq \sup_{\bs{\Lambda}\in \mathbb{R}^{n \times R}} \left\{ \frac{1}{n} \log (\det(\bs{S}(\bs{\rho}))) - \frac{1}{2}\log\left( \hat{\sigma}^{2}(\bs{\theta},\bs{\Lambda})\right) \right\}. \label{P2.1} 
\end{IEEEeqnarray}
Evaluated at $\bs{\theta}^0$, a lower bound, denoted $\underline{\mathcal{L}}(\bs{\theta}^0)$, can be established by substituting in the true DGP, and using Assumptions \ref{AS:1.1} and \ref{AS:1.2}, 
\begin{align}
\underline{\mathcal{L}}(\bs{\theta}^0) 
&\coloneqq \frac{1}{n} \log (\det(\bs{S}))-   \frac{1}{2}\log\left( \sigma^2_0 +O_P \left( {\frac{1}{\min\{n,T\}}} \right) \right) \notag \\
&= \frac{1}{n} \log (\det(\bs{S}))-   \frac{1}{2}\log\left( \sigma^2_0 + O_P ( a_{nT}^2 ) \right) 
\leq \mathcal{L}(\bs{\theta}^0).  \label{tttt}
\end{align}
Second, using Lemmas \ref{L1.2}, \ref{L1.12}, \ref{L1.15}, \ref{L1.18} and Assumption \ref{new1}, an upper bound for $\mathcal{L}(\bs{{\theta}})$, denoted $\bar{\mathcal{L}}(\bs{{\theta}})$, can also be established,
\begin{align}
\mathcal{L}(\bs{{\theta}}) 
\leq&\
\frac{1}{n}\log(\det(\bs{S}(\bs{{\rho}}))) - \frac{1}{2} \log \Bigg( c_1 ||\bs{\theta}^0 - \bs{{\theta}}||_2^2  + O_P \left( \frac{1}{\min\{n,T\}} \right)  \notag \\
&+ 
\frac{\sigma^2_0}{n}\text{tr}((\bs{S}(\bs{{\rho}})\bs{S}^{-1} )'\bs{S}(\bs{{\rho}}) \bs{S}^{-1}) + O_P\left(\frac{1}{\sqrt{nT}}\right) + ||\bs{\theta}^0 - \bs{{\theta}}||_2 O_P\left( \sqrt{\frac{P}{nT}} \right) \Bigg) \notag \\
=&\ 
\frac{1}{n}\log(\det(\bs{S}(\bs{{\rho}}))) - \frac{1}{2} \log \Big(  c_1 ||\bs{{\theta}} - \bs{\theta}^0||_2^2 + O_P(a_{nT}) ||\bs{{\theta}} - \bs{\theta}^0||_2 + O_P(a_{nT}^2)  \notag   \\
&+ \frac{\sigma^2_0}{n}\text{tr}((\bs{S}(\bs{{\rho}})\bs{S}^{-1} )'\bs{S}(\bs{{\rho}}) \bs{S}^{-1})  \Big) 
\notag \\
\eqqcolon&\ \bar{\mathcal{L}}(\bs{{\theta}}). \label{P2.20}
\end{align}
Now, since $\bs{\tilde{\theta}}$ is a global maximiser, $\mathcal{L}(\bs{\theta}^0) \leq \mathcal{L}(\bs{\tilde{\theta}})$ and therefore $\underline{\mathcal{L}}(\bs{\theta}^0) \leq \bar{\mathcal{L}}(\bs{\tilde{\theta}})$. Using the expressions for these bounds derived in \eqref{tttt} and \eqref{P2.20} gives
\begin{IEEEeqnarray}{rCl}
\IEEEeqnarraymulticol{3}{l}{\frac{1}{n}\log(\det(\bs{S})) - \frac{1}{2} \log\left(\sigma^2_0 +O_P (a_{nT}^2 ) \right)} \notag \\
&\leq& \frac{1}{n}\log(\det(\bs{S}(\bs{\tilde{\rho}}))) - \frac{1}{2} \log \Big(  c_1 ||\bs{\tilde{\theta}} - \bs{\theta}^0||_2^2 + O_P(a_{nT}) ||\bs{\tilde{\theta}} - \bs{\theta}^0||_2 + O_P(a_{nT}^2)    \notag \\
&&+ \frac{\sigma^2_0}{n}\text{tr}((\bs{S}(\bs{\tilde{\rho}})\bs{S}^{-1} )'\bs{S}(\bs{\tilde{\rho}}) \bs{S}^{-1}) \Big). \label{ADED3}
\end{IEEEeqnarray}
Multiplying both sides of \eqref{ADED3} by $-2$, exponentiating, and then noticing that, by Lemma \hyperref[firstlem]{A.1}, $\sigma^2_0 \text{det}((\bs{S}(\bs{\tilde{\rho}})\bs{S}^{-1} )'\bs{S}(\bs{\tilde{\rho}}) \bs{S}^{-1})^{\frac{1}{n}} \leq \frac{\sigma^2_0}{n}\text{tr}((\bs{S}(\bs{\tilde{\rho}})\bs{S}^{-1} )'\bs{S}(\bs{\tilde{\rho}}) \bs{S}^{-1})$, results in 
\begin{align}
0
&\geq c_1 ||\bs{\tilde{\theta}} - \bs{\theta}^0||_2^2 + O_P(a_{nT}) ||\bs{\tilde{\theta}} - \bs{\theta}^0||_2 + O_P(a_{nT}^2). \label{ADED5}
\end{align}
Completing the square, $0 \geq (\sqrt{c_1} ||\bs{\tilde{\theta}} - \bs{\theta}^0||_2 + O_P(a_{nT}) )^2  + O_P(a_{nT}^2)$, whereby it follows that $||\bs{\tilde{\theta}} - \bs{\theta}^0||_2 = O_P(a_{nT})$. \\

\noindent \textbf{Consistency of the PQMLE $\bs{\hat{\theta}}$}\\
Since $\bs{\hat{\theta}}$ is the maximiser of the penalised quasi-likelihood function, $\mathcal{Q}({\bs{\theta}^0}) \leq \mathcal{Q}(\bs{\hat{\theta}})$. Thus,
\begin{align}
\mathcal{Q}({\bs{\theta}^0})
&=
\mathcal{L}({\bs{\theta}^0}) 
- 
\left( \gamma_{\rho} \sum^{Q}_{q=1} \omega_{q} |\rho_{q}^0|
+
\gamma_{\beta} \sum^{K}_{k=1} \omega_{Q+k} |\beta_{k}^0| \right) \notag \\
&\leq 
\mathcal{Q}(\bs{\hat{\theta}}) \notag \\
&=
\mathcal{L}(\bs{\hat{\theta}})
-
\left( \gamma_{\rho} \sum^{Q}_{q=1} \omega_{q} |\hat{\rho}_{q}|
+
\gamma_{\beta} \sum^{K}_{k=1} \omega_{Q+k} |\hat{\beta}_{k}| \right) \notag \\
&\leq 
\mathcal{L}(\bs{\hat{\theta}}).
\label{P2.23}
\end{align}
Consider the penalty term. Under Assumption \ref{AS:4.0},
\begin{align}
\gamma_{\rho} \sum^{Q}_{q=1} \omega_{q} |\rho_{q}^0|
+
\gamma_{\beta} \sum^{K}_{k=1} \omega_{Q+k} |\beta_{k}^0| 
&\leq
c_2 \max\{\gamma_{\rho},\gamma_{\beta}\} P^0 \left( \left|  \frac{{\theta}_{\underline{p}}^{\dagger}}{\theta^0_{\underline{p}}} \right| \right)^{-\zeta} |\theta^0_{\underline{p}}|^{-\zeta}, 
\end{align}
where $\underline{p} \coloneqq \argmin_{1 \leq p \leq P:\color{white}.\color{black}\theta^0_p \neq 0}
|{\theta}_{{p}}^{\dagger}|$. Since the initial estimate ${\bs{\theta}}^{\dagger}$ satisfies $||{\bs{\theta}}^{\dagger} - \bs{\theta}^0||_2 = O_P(r_{nT}) = o_P(1)$, it follows that
$|{{\theta}_{\underline{p}}^{\dagger}}/{{\theta}}^0_{\underline{p}}- 1 | \leq \frac{1}{|{\theta}^0_{\underline{p}}|}||{\bs{\theta}}^{\dagger}  - \bs{\theta}^0||_2 = o_P(1) $
which implies ${{\theta}_{\underline{p}}^{\dagger}} \slash {{\theta}}^0_{\underline{p}} = O_P(1)$. Hence, 
\begin{align}
\gamma_{\rho} \sum^{Q}_{q=1} \omega_{q} |\rho_{q}^0|
+
\gamma_{\beta} \sum^{K}_{k=1} \omega_{Q+k} |\beta_{k}^0| 
=
\max\{\gamma_{\rho},\gamma_{\beta}\} O_P(P^0) 
=
O_P(a_{nT}^2), \label{thiseq}
\end{align}
under Assumption \ref{AS:4.1}. Next, using \eqref{thiseq}, and applying the lower and upper bounds derived in \eqref{tttt} and \eqref{P2.20} to \eqref{P2.23} gives
\begin{align}
&\frac{1}{n}\log(\det(\bs{S})) - \frac{1}{2} \log(\sigma^2_0 + O_P(a_{nT}^2))
+ O_P(a_{nT}^2) \leq
\frac{1}{n}\log(\det(\bs{S}(\bs{\hat{\rho}}))) \notag \\
&- \frac{1}{2} \log \Big(  c_1 ||\bs{\hat{\theta}} - \bs{\theta}^0||_2^2 + O_P(a_{nT}) ||\bs{\hat{\theta}} - \bs{\theta}^0||_2 + O_P(a_{nT}^2)  + \frac{\sigma^2_0}{n}\text{tr}((\bs{S}(\bs{\hat{\rho}})\bs{S}^{-1} )'\bs{S}(\bs{\hat{\rho}}) \bs{S}^{-1})  \Big).
\end{align}
After rearranging and simplifying this becomes  
\begin{align}
&\log\left( \sigma^2_0 \text{det}((\bs{S}(\bs{\hat{\rho}})\bs{S}^{-1} )'\bs{S}(\bs{\hat{\rho}}) \bs{S}^{-1})^{\frac{1}{n}} + O_P(a_{nT}^2)\right) 
+ O_P(a_{nT}^2) \notag \\
&\geq
\log \Big(  c_1 ||\bs{\hat{\theta}} - \bs{\theta}^0||_2^2 + O_P(a_{nT}) ||\bs{\hat{\theta}} - \bs{\theta}^0||_2 + O_P(a_{nT}^2) + \frac{\sigma^2_0}{n}\text{tr}((\bs{S}(\bs{\hat{\rho}})\bs{S}^{-1} )'\bs{S}(\bs{\hat{\rho}}) \bs{S}^{-1})  \Big).
\end{align}
Exponentiating, using Lemma \hyperref[firstlem]{A.1}, and the fact that by Assumption \ref{AS:6.3} $O_P(a_{nT}^2) = o_P(1)$ gives the result
\begin{align}
0 \geq c_1 ||\bs{\hat{\theta}} - \bs{\theta}^0||_2^2 + O_P(a_{nT}) ||\bs{\hat{\theta}} - \bs{\theta}^0||_2 + O_P(a_{nT}^2), \label{samething}
\end{align}
whereby completing the square yields $||\bs{\hat{\theta}} - \bs{\theta}^0||_2 = O_P(a_{nT})$.  
\end{pff}

\begin{pff}[Proof of Proposition \ref{Prop3}] 
Since the PQMLE $\bs{\hat{\theta}}$ is consistent for $\bs{\theta}^0$ by Proposition \ref{Prop2}, and by Assumption \ref{AS:2.2} $\bs{\theta}^0$ is in the interior of
$\bs{\Theta}$, $\bs{\hat{\theta}}$ must also be in the interior of $\bs{\Theta}$ w.p.a.1 as $n,T \rightarrow \infty$. Thus, w.p.a.1, $\bs{\hat{\theta}}$ must solve the first order condition
\begin{align}
\frac{\partial \mathcal{Q}(\bs{{\theta}}, \bs{\Lambda})}{\partial \bs{\theta}} = \frac{\partial \mathcal{L}(\bs{{\theta}}, \bs{\Lambda})}{\partial \bs{\theta}} - \frac{\partial {\varrho} (\bs{\theta}, \bs{\gamma}, \zeta)}{\partial \bs{\theta}}  = \bs{0}_{P \times 1}, \label{P3.1}
\end{align}
where
\begin{align}
\frac{\partial \mathcal{L}(\bs{{\theta}}, \bs{\Lambda})}{\partial \bs{\theta}} 
& =
\begin{pmatrix}
  -\frac{1}{n}\text{tr}(\bs{G}_1(\bs{{\rho}})) + \frac{1}{{\hat{\sigma}^2}(\bs{{\theta}}, \bs{\Lambda})} \frac{1}{nT} \sum^{T}_{t=1} (\bs{W}_1 \bs{y}_t)' \bs{M}_{\bs{\Lambda}} (\bs{S}(\bs{{\rho}}) \bs{y}_t - \bs{{X}}_t {\bs{\beta}}) \\
  \vdots \\
  -\frac{1}{n}\text{tr}(\bs{G}_Q(\bs{{\rho}})) + \frac{1}{{\hat{\sigma}^2}(\bs{{\theta}}, \bs{\Lambda})} \frac{1}{nT} \sum^{T}_{t=1} (\bs{W}_{\kern -0.2em Q} \bs{y}_t)' \bs{M}_{\bs{\Lambda}} (\bs{S}(\bs{{\rho}}) \bs{y}_t - \bs{{X}}_t {\bs{\beta}}) \\
  \frac{1}{{\hat{\sigma}^2}(\bs{{\theta}}, \bs{\Lambda})} {\frac{1}{nT} \sum^{T}_{t=1} \bs{x}_{1t}' \bs{M}_{\bs{\Lambda}} (\bs{S}(\bs{{\rho}}) \bs{y}_t - \bs{{X}}_t {\bs{\beta}})} \\
  \vdots \\
  \frac{1}{{\hat{\sigma}^2}(\bs{{\theta}}, \bs{\Lambda})} {\frac{1}{nT} \sum^{T}_{t=1} \bs{x}_{Kt}' \bs{M}_{\bs{\Lambda}} (\bs{S}(\bs{{\rho}}) \bs{y}_t - \bs{{X}}_t {\bs{\beta}})}
\end{pmatrix}. \label{P3.2}
\end{align}
In the following it is shown that, as $n,T \rightarrow \infty$, this first order condition cannot hold unless the estimators of those ${\theta}_p$ which have a true value of zero also take a value of exactly zero w.p.a.1. To reach a contradiction, suppose that there is some $p$, call this $p^*$, for which $\theta^0_p = 0$ yet $\text{Pr}(\hat{\theta}_p = 0)$ does not go to 1 as $n,T \rightarrow \infty$. It is first shown that $\frac{\partial \mathcal{L}(\bs{{\theta}}, \bs{\Lambda})}{\partial {\theta}_{p^*}} |_{\bs{\theta} = \bs{\hat{\theta}}} = O_P(1)$, i.e., the first order condition evaluated at $\bs{\hat{\theta}}$ is not explosive in probability. Since ${\theta}_{p^*}$ could be some $\rho_q$ or $\beta_k$, both cases are examined in turn. Consider first the case where $\theta_{p^*}$ is some $\rho_q$. Substituting in the true data generating process, the element of $\frac{\partial \mathcal{L}(\bs{{\theta}}, \bs{\Lambda})}{\partial \bs{\theta}} |_{\bs{\theta} = \bs{\hat{\theta}}}$ relating to $\rho_q$ is equal to
\begin{align}
&-\frac{1}{n}\text{tr}(\bs{G}_q(\bs{\hat{\rho}})) + \frac{1}{{\hat{\sigma}^2}(\bs{\hat{\theta}}, \bs{\Lambda})} \frac{1}{nT} \sum^{T}_{t=1} (\bs{W}_{\kern -0.2em q} \bs{y}_t)' \bs{M}_{\bs{\Lambda}} (\bs{S}(\bs{\hat{\rho}}) \bs{y}_t - \bs{{X}}_t \bs{\hat{\beta}}) \notag \\
=& -\frac{1}{n}\text{tr}(\bs{G}_q (\bs{\hat{\rho}})) 
+ \frac{1}{{\hat{\sigma}^2}(\bs{\hat{\theta}}, \bs{\Lambda})} \frac{1}{nT} \sum^{T}_{t=1} (\bs{G}_q \bs{{X}}_t \bs{\beta}^0)' \bs{M}_{\bs{\Lambda}} \bs{Z}_t(\bs{\theta}^0 - \bs{\hat{\theta}}) \notag \\
&+ \frac{1}{{\hat{\sigma}^2}(\bs{\hat{\theta}}, \bs{\Lambda})} \frac{1}{nT} \sum^{T}_{t=1} (\bs{G}_q \bs{{X}}_t \bs{\beta}^0)' \bs{M}_{\bs{\Lambda}} \bs{S}(\bs{\hat{\rho}}) \bs{S}^{-1} \bs{\Lambda}^0\bs{f}^0_t 
+ \frac{1}{{\hat{\sigma}^2}(\bs{\hat{\theta}}, \bs{\Lambda})} \frac{1}{nT} \sum^{T}_{t=1} (\bs{G}_q  \bs{{X}}_t \bs{\beta}^0)' \bs{M}_{\bs{\Lambda}} \bs{S}(\bs{\hat{\rho}}) \bs{S}^{-1} \bs{\varepsilon}_t \notag \\
&+ \frac{1}{{\hat{\sigma}^2}(\bs{\hat{\theta}}, \bs{\Lambda})} \frac{1}{nT} \sum^{T}_{t=1} (\bs{G}_q  \bs{\Lambda}^0 \bs{f}^0_t)' \bs{M}_{\bs{\Lambda}} \bs{Z}_t(\bs{\theta}^0 - \bs{\hat{\theta}}) 
+ \frac{1}{{\hat{\sigma}^2}(\bs{\hat{\theta}}, \bs{\Lambda})} \frac{1}{nT} \sum^{T}_{t=1} (\bs{G}_q \bs{\Lambda}^0\bs{f}^0_t)' \bs{M}_{\bs{\Lambda}} \bs{S}(\bs{\hat{\rho}}) \bs{S}^{-1}\bs{\Lambda}^0\bs{f}^0_t \notag \\
&+ \frac{1}{{\hat{\sigma}^2}(\bs{\hat{\theta}}, \bs{\Lambda})} \frac{1}{nT} \sum^{T}_{t=1} (\bs{G}_q  \bs{\Lambda}^0 \bs{f}^0_t)' \bs{M}_{\bs{\Lambda}} \bs{S}(\bs{\hat{\rho}}) \bs{S}^{-1} \bs{\varepsilon}_t 
+ \frac{1}{{\hat{\sigma}^2}(\bs{\hat{\theta}}, \bs{\Lambda})} \frac{1}{nT} \sum^{T}_{t=1} (\bs{G}_q  \bs{\varepsilon}_t)' \bs{M}_{\bs{\Lambda}} \bs{Z}_t(\bs{\theta}^0 - \bs{\hat{\theta}}) \notag \\
&+\frac{1}{{\hat{\sigma}^2}(\bs{\hat{\theta}}, \bs{\Lambda})} \frac{1}{nT} \sum^{T}_{t=1} (\bs{G}_q  \bs{\varepsilon}_t)' \bs{M}_{\bs{\Lambda}} \bs{S}(\bs{\hat{\rho}}) \bs{S}^{-1}\bs{\Lambda}^0\bs{f}^0_t
+ \frac{1}{{\hat{\sigma}^2}(\bs{\hat{\theta}}, \bs{\Lambda})} \frac{1}{nT} \sum^{T}_{t=1} (\bs{G}_q  \bs{\varepsilon}_t)' \bs{M}_{\bs{\Lambda}} \bs{S}(\bs{\hat{\rho}}) \bs{S}^{-1}\bs{\varepsilon}_t \notag \\
\eqqcolon&\ s_{1} + \ldots + s_{10}. \label{P3.16}
\end{align}
Since $\bs{G}_q(\bs{\rho})$ is UB, terms ${s_{5}},\ldots,s_{10}$ are $O_P(1)$ by the same arguments as for their counterparts in the proof of Lemma \ref{L1.19} (terms $l_{2},\ldots,l_{6}$; see Supplementary Material), and using the result in that lemma (whereby $1/{\hat{\sigma}^2}(\bs{\hat{\theta}}, \bs{\Lambda}) = O_P(1)$). Since the rank of $\bs{G}_q(\bs{{\rho}})$ can be no greater than $n$, using $|\text{tr}(\bs{B})| \leq \text{rank}(\bs{B})||\bs{B}||_2$ for some square matrix $\bs{B}$ \citep[][Lemma S.4.1(v)]{moon_dynamic_nodate}, and that $\bs{S}^{-1}(\bs{{\rho}})$ and $\bs{W}_{\kern -0.2em q}$ are UB, one has
\begin{IEEEeqnarray}{rCl}
|s_{1}| 
&=& \frac{1}{n}|\text{tr}(\bs{G}_q(\bs{\hat{\rho}}))| \leq ||\bs{G}_q(\bs{\hat{\rho}})||_2 \leq ||\bs{S}^{-1}(\bs{\hat{\rho}})||_2 ||\bs{W}_{\kern -0.2em q}||_2= O_P(1).\color{white}-\color{black}
\end{IEEEeqnarray}
Using Lemmas \ref{L1.17}, \ref{L1.18} and \ref{L1.19}, as well as Proposition \ref{Prop2}, yields 
\begin{align}
|s_{2}| 
&\leq  \frac{1}{\sqrt{nT}}\frac{1}{{\hat{\sigma}^2}(\bs{\hat{\theta}}, \bs{\Lambda})} ||\bs{G}_q||_2 ||\bs{M}_{\bs{\Lambda}}||_2    \left( \sum^{T}_{t=1} ||\bs{{X}}_t\bs{\beta}^0||_2^2 \right)^{\frac{1}{2}} \left( \frac{1}{nT} \sum^{T}_{t=1} ||\bs{Z}_t(\bs{\theta}^0 - \bs{\hat{\theta}})||_2^2 \right)^{\frac{1}{2}} \notag \\
&= \frac{1}{\sqrt{nT}}   O_P(\sqrt{nT}) O_P(a_{nT}) = O_P(1). \label{P3.17}
\end{align}
The remaining terms, $s_{3}$ and $s_{4}$, can be shown to be $O_P(1)$ similarly, using Lemmas \ref{L1.10}, \ref{L1.16} \ref{L1.17} and \ref{L1.19}. Next consider the case where $\theta_{p^*}$ is some $\beta_k$. The element of $\frac{\partial \mathcal{L}(\bs{{\theta}}, \bs{\Lambda})}{\partial \bs{\theta}} |_{\bs{\theta} = \bs{\hat{\theta}}}$ corresponding to $\beta_k$ is
\begin{IEEEeqnarray}{rCl}
\IEEEeqnarraymulticol{3}{l}{
\frac{1}{{\hat{\sigma}^2}(\bs{\hat{\theta}}, \bs{\Lambda})} \frac{1}{nT} \sum^T_{t=1} \bs{x}'_{kt} \bs{M}_{\bs{\Lambda}} (\bs{S}(\bs{\hat{\rho}})\bs{S}^{-1}(\bs{{X}}_t\bs{\beta}^0 + \bs{\Lambda}^0\bs{f}^0_t + \bs{\varepsilon}_t) - \bs{{X}}_t \bs{\hat{\beta}})} \notag \\
&=& \frac{1}{{\hat{\sigma}^2}(\bs{\hat{\theta}}, \bs{\Lambda})} \frac{1}{nT} \sum^T_{t=1} \bs{x}'_{kt} \bs{M}_{\bs{\Lambda}} \bs{Z}_t(\bs{\theta}^0 - \bs{\theta}) 
+ \frac{1}{{\hat{\sigma}^2}(\bs{\hat{\theta}}, \bs{\Lambda})} \frac{1}{nT} \sum^T_{t=1} \bs{x}'_{kt} \bs{M}_{\bs{\Lambda}} \bs{S}(\bs{\hat{\rho}})\bs{S}^{-1}\bs{\Lambda}^0 \bs{f}^0_t \notag \\
&& +\ \frac{1}{{\hat{\sigma}^2}(\bs{\hat{\theta}}, \bs{\Lambda})} \frac{1}{nT} \sum^T_{t=1} \bs{x}'_{kt} \bs{M}_{\bs{\Lambda}} \bs{S}(\bs{\hat{\rho}})\bs{S}^{-1}\bs{\varepsilon}_t \notag \\ 
&\eqqcolon& p_{1} + p_{2} + p_{3}. \notag
\end{IEEEeqnarray}
Using Lemmas \ref{L1.9}, \ref{L1.10}, \ref{L1.16}, \ref{L1.18} and \ref{L1.19}, one has
\begin{align}
|p_{1}|
&\leq  \frac{1}{nT} \frac{1}{{\hat{\sigma}^2}(\bs{\hat{\theta}}, \bs{\Lambda})} ||\bs{M}_{\bs{\Lambda}}||_2  \left( \sum^T_{t=1} ||\bs{x}_{kt}||_2^2 \right)^{\frac{1}{2}}  \left( \sum^T_{t=1} ||\bs{Z}_t(\bs{\theta}^0 - \bs{\hat{\theta}})||_2^2 \right)^{\frac{1}{2}}       \notag \\
&= \frac{1}{\sqrt{nT}} \frac{1}{{\hat{\sigma}^2}(\bs{\hat{\theta}}, \bs{\Lambda})} ||\bs{M}_{\bs{\Lambda}}||_2  ||\bs{\mathcal{X}}_{k}||_F  \left(\frac{1}{nT} \sum^T_{t=1} ||\bs{Z}_t(\bs{\theta}^0 - \bs{\hat{\theta}})||_2^2 \right)^{\frac{1}{2}}   \notag \\
&= \frac{1}{\sqrt{nT}} O_P(\sqrt{nT}) O_P(a_{nT}) = O_P(1), \label{P3.13}
\end{align}
\begin{align}
|p_{2}| 
&\leq \frac{1}{nT}  \frac{1}{{\hat{\sigma}^2}(\bs{\hat{\theta}}, \bs{\Lambda})} ||\bs{M}_{\bs{\Lambda}}||_2 ||\bs{S}(\bs{\hat{\rho}})\bs{S}^{-1}||_2 ||\bs{\Lambda}^0||_2 \left( \sum^T_{t=1} ||\bs{x}_{kt}||_2^2 \right)^{\frac{1}{2}} \left( \sum^T_{t=1}||\bs{f}^0_t||_2^2 \right)^{\frac{1}{2}} \notag \\
&= \frac{1}{nT}  \frac{1}{{\hat{\sigma}^2}(\bs{\hat{\theta}}, \bs{\Lambda})} ||\bs{M}_{\bs{\Lambda}}||_2 ||\bs{S}(\bs{\hat{\rho}})\bs{S}^{-1}||_2 ||\bs{\Lambda}^0||_2 ||\bs{\mathcal{X}}_{k}||_F ||\bs{F}^0||_F \notag \\
&= \frac{1}{nT} O_P(\sqrt{n}) O_P(\sqrt{T}) O_P(\sqrt{nT}) = O_P(1), \label{P3.14}
\end{align}
and
\begin{align}
|p_{3}|
&\leq \frac{1}{nT}  \frac{1}{{\hat{\sigma}^2}(\bs{\hat{\theta}}, \bs{\Lambda})} ||\bs{M}_{\bs{\Lambda}}||_2 ||\bs{S}(\bs{\hat{\rho}})\bs{S}^{-1}||_2  \left( \sum^T_{t=1} ||\bs{x}_{kt}||_2^2 \right)^{\frac{1}{2}} \left( \sum^T_{t=1}||\bs{\varepsilon}_t||_2^2 \right)^{\frac{1}{2}} \notag \\
&= \frac{1}{nT}  \frac{1}{{\hat{\sigma}^2}(\bs{\hat{\theta}}, \bs{\Lambda})} ||\bs{M}_{\bs{\Lambda}}||_2 ||\bs{S}(\bs{\hat{\rho}})\bs{S}^{-1}||_2  ||\bs{\mathcal{X}}_{k}||_F ||\bs{\varepsilon}||_F \notag \\
&= \frac{1}{nT} O_P(\sqrt{nT}) O_P(\sqrt{nT}) = O_P(1). \label{P3.15}
\end{align}
Combining the previous results gives $\frac{\partial \mathcal{L}(\bs{{\theta}}, \bs{\Lambda})}{\partial \theta_{p^*}} |_{\bs{\theta} = \bs{\hat{\theta}}} = O_P(1)$. Turning now to the derivative of the penalty term, evaluated at $\bs{\hat{\theta}}$,
\begin{align}
\frac{\partial \varrho(\bs{{\theta}}, \bs{\gamma},{\zeta})}{\partial \theta_{p^*}} \Big|_{\bs{\theta} 
	= \bs{\hat{\theta}}}
=
 - \gamma^{*} \frac{1}{|{\theta}_{{p^*}}^{\dagger}|^{\zeta}} \frac{\hat{\theta}_{{p^*}}}{|\hat{\theta}_{{p^*}}|}, 
\end{align}
where $\gamma^{*} \in \{\gamma_{\rho},\gamma_{\beta}\}$ denotes the penalty parameter associated with ${\theta}_{p^*}$.  By Assumption \ref{ASS6}, $\min\{\gamma_{\rho} , \gamma_{\beta}\} |{\theta}_{p^*}^{\dagger}|^{-\zeta}$ is explosive in probability because ${\theta}_{p^*}^0 = 0$ and so $|{\theta}_{p^*}^{\dagger}| = |{\theta}_{p^*}^{\dagger} - {\theta}_{p^*}^0| \leq ||\bs{\theta}^{\dagger} - \bs{\theta}^0||_2 = o_P(1)$ by Assumption \ref{AS:4.2}. As such, as $n,T \rightarrow \infty$, the first order condition cannot be satisfied since $\frac{\partial \mathcal{L}(\bs{{\theta}}, \bs{\Lambda})}{\partial {\theta}_{p^*}} |_{\bs{\theta} = \bs{\hat{\theta}}} = O_P(1)$ and yet the derivative of the penalty term diverges. This contradicts $\bs{\hat{\theta}}$ being a maximiser of the objective function. Therefore, instead, it must be that $\hat{\theta}_{p^*} = 0$ w.p.a.1 as $n,T \rightarrow \infty$ for the first order condition \eqref{P3.1} to be satisfied. 
\end{pff}

\begin{pff}[Proof of Theorem \ref{Thrm1}] 
Starting with the expression obtained in Lemma \hyperref[Prop4]{{A.4}}, 
\begin{align}
\bs{D}
\sqrt{nT} (\bs{\hat{\theta}}_{(1)} - \bs{{\theta}}_{(1)}^0)
=&\ \frac{1}{\sigma^2_0} \frac{1}{\sqrt{nT} }\bs{\mathcal{Z}}_{(1)}' (\bs{M}_{\bs{F}^0} \otimes \bs{M}_{\bs{\Lambda}^0}) \textup{vec}(\bs{\varepsilon}) \notag \\
&+\ \frac{1}{\sigma^2_0} \frac{1}{\sqrt{nT}}
\begin{pmatrix}
\text{tr} \left( (\bs{G}_1^* \bs{\varepsilon} )' \bs{M}_{\bs{\Lambda}^0} \bs{\varepsilon} \bs{M}_{\bs{F}^0} \right) \\
\vdots \\
\text{tr} \left ( (\bs{G}_{Q^0}^*  \bs{\varepsilon} )' \bs{M}_{\bs{\Lambda}^0} \bs{\varepsilon} \bs{M}_{\bs{F}^0} \right) \\
\bs{0}_{K^0 \times 1} 
\end{pmatrix} 
+ \bs{o}_P(1) \notag \\
=&\ 
\frac{1}{\sigma^2_0} \frac{1}{\sqrt{nT} } \bs{c} 
-\frac{1}{\sigma^2_0} \frac{1}{\sqrt{nT} }
\begin{pmatrix}
\text{tr} \left( (\bs{G}_1^* \bs{\varepsilon} )' ( \bs{P}_{\bs{\Lambda}^0} \bs{\varepsilon} + \bs{M}_{\bs{\Lambda}^0}  \bs{\varepsilon} \bs{P}_{\bs{F}^0} ) \right) \\
\vdots \\
\text{tr} \left ( (\bs{G}_{Q^0}^* \bs{\varepsilon} )' ( \bs{P}_{\bs{\Lambda}^0} \bs{\varepsilon} + \bs{M}_{\bs{\Lambda}^0}  \bs{\varepsilon} \bs{P}_{\bs{F}^0} )\right) \\
\bs{0}_{K^0 \times 1} 
\end{pmatrix} 
 \notag \\
&-\frac{1}{\sigma^2_0} \frac{1}{\sqrt{nT} }
\begin{pmatrix}
\text{tr} \left( ( \bs{\mathscr{Z}}_1 - \bs{\bar{\mathscr{Z}}}_1 )' ( \bs{P}_{\bs{\Lambda}^0} \bs{\varepsilon} + \bs{M}_{\bs{\Lambda}^0}  \bs{\varepsilon} \bs{P}_{\bs{F}^0} ) \right) \\
\vdots \\
\text{tr} \left ( ( \bs{\mathscr{Z}}_{P^0} - \bs{\bar{\mathscr{Z}}}_{P^0} )' ( \bs{P}_{\bs{\Lambda}^0} \bs{\varepsilon} + \bs{M}_{\bs{\Lambda}^0}  \bs{\varepsilon} \bs{P}_{\bs{F}^0} )\right) 
\end{pmatrix}
+ \bs{o}_P(1), \label{rrrr}
\end{align}
where $\bs{c} \coloneqq \pmb{\mathbb{Z}}_{(1)}' \textup{vec}(\bs{\varepsilon})
+ 
(\text{tr}(\bs{\varepsilon}'\bs{G}^*_{1}\bs{\varepsilon}),\ldots,
\text{tr}(\bs{\varepsilon}'\bs{G}^*_{Q^0}\bs{\varepsilon}),\bs{0}_{1 \times K^0})'$, recalling the definition of $\pmb{\mathbb{Z}}_{(1)}$ given just prior to the statement of Assumption \ref{ASS8}. By expanding the second term on the right-hand side of \eqref{rrrr} and applying Lemmas \ref{aaadd3}, \ref{aaadd4}, \ref{aaadd5}, and also applying Lemma \ref{aaadd6} to the third term, one obtains   
\begin{align}
\bs{D}
\sqrt{nT} (\bs{\hat{\theta}}_{(1)} - \bs{{\theta}}_{(1)}^0) 
&= \frac{1}{\sigma^2_0} \frac{1}{\sqrt{nT} } \bs{c} 
+  \pmb{\mathbbm{b}} + \bs{o}_P(1).
\end{align}	
Rearranging and premultiplying by $\big(\pmb{\mathbb{S}}({\bs{D}} + {\bs{V}})\pmb{\mathbb{S}}'\big)^{-\frac{1}{2}}$ gives
\begin{align}
	\sqrt{nT}\big(\pmb{\mathbb{S}}({\bs{D}} + {\bs{V}})\pmb{\mathbb{S}}'\big)^{-\frac{1}{2}}\pmb{\mathbb{S}} \bs{D} ( \bs{\hat{\theta}}_{(1)}  - \bs{D}^{-1} \pmb{\mathbbm{b}} - \bs{\theta}^0_{(1)}) 
	= \big(\pmb{\mathbb{S}}({\bs{D}} + {\bs{V}})\pmb{\mathbb{S}}'\big)^{-\frac{1}{2}}\pmb{\mathbb{S}} \frac{1}{\sqrt{nT}} \frac{1}{\sigma^2_0} \bs{c}+ \bs{o}_P(1). \label{TR1-5}
\end{align}
Finally, using Lemma \hyperref[aaadd2]{A.6} and Assumption \ref{AS:10.4}, $\big(\pmb{\mathbb{S}}({\bs{D}} + {\bs{V}})\pmb{\mathbb{S}}'\big)^{-\frac{1}{2}}\pmb{\mathbb{S}} \frac{1}{\sqrt{nT}} \frac{1}{\sigma^2_0} \bs{c} \xrightarrow{d} \mathcal{N}(\bs{0}_{L\times 1}, \bs{I}_{L})$, which yields the result. 
\end{pff}

\begin{pff}[Proof of Proposition \ref{Prop5}] 
In order to prove the result, it suffices to show that $
||\bs{D}^{-1} \pmb{{\mathbbm{b}}} - \bs{\hat{D}}^{-1}  \pmb{\hat{\mathbbm{b}}}||_2 = o_P(1)$.
Observe that
\begin{align}
||\bs{D}^{-1} \pmb{{\mathbbm{b}}} - \bs{\hat{D}}^{-1}  \pmb{\hat{\mathbbm{b}}}||_2
&\leq ||\bs{D}^{-1} - \bs{\hat{D}}^{-1}||_2 ||\pmb{\hat{\mathbbm{b}}}||_2 + ||\bs{D}^{-1}||_2 ||\pmb{{\mathbbm{b}}} -  \pmb{\hat{\mathbbm{b}}}||_2. \label{ven}
\end{align}
It is straightforward to establish that $||\bs{D}^{-1} - \bs{\hat{D}}^{-1}||_2 ||\pmb{\hat{\mathbbm{b}}}||_2 = \bs{o}_P(1)$ using Lemma \ref{sub11} and the fact that, under Assumptions  \ref{ASS1}--\ref{ASS7}, $||\bs{\hat{\theta}} - \bs{\theta}^0||_2 = O_P\left( \sqrt{\frac{P}{nT}} \right)$, which follows from (F.17) in the proof of Lemma {\hyperref[Prop4]{\text{A.4}}} in the Supplementary Material. For the second term in \eqref{ven}, $||\pmb{{\mathbbm{b}}} -  \pmb{\hat{\mathbbm{b}}}||_2 = o_P\left( 1 \right)$ can be shown using Lemmas \ref{aaadd3}--\ref{aaadd6}, and the following two results. To simplify notation, assume that $P = P^0$, $Q = Q^0$ and $\phi_1^0$ is nonzero. First,
\begin{align}
||\bs{G}^*_q - \bs{G}_q^*(\bs{\hat{\rho}})||_2 
&= ||\bs{G}_q - \bs{G}_q(\bs{\hat{\rho}}) - \frac{1}{n}\text{tr}(\bs{G}_q)\bs{I}_n  + \frac{1}{n} \text{tr}(\bs{G}_q(\bs{\hat{\rho}}))\bs{I}_n ||_2 \notag \\
&\leq ||\bs{G}_q - \bs{G}_q(\bs{\hat{\rho}})||_2 + \frac{1}{n} |\text{tr}(\bs{G}_q(\bs{\hat{\rho}})-\bs{G}_q)| \notag \\
&\leq 2||\bs{G}_q - \bs{G}_q(\bs{\hat{\rho}})||_2 \notag \\
&= 2||\bs{G}_q(\bs{\hat{\rho}}) (\bs{S}(\bs{\hat{\rho}})\bs{S}^{-1} - \bs{I}_n)||_2 \notag \\
&\leq 2||\bs{G}_q(\bs{\hat{\rho}})||_2 ||\bs{S}(\bs{\hat{\rho}})\bs{S}^{-1} - \bs{I}_n||_2 \notag \\
&= O_P(\sqrt{{Q}}||\bs{\theta}^0 - \bs{\hat{\theta}}||_2),
\end{align}
using Lemma \ref{L2.1}. Second, 
\begin{align}
	||\bs{A} - \bs{A}(\bs{\hat{\rho}},\bs{\hat{\phi}})||_2 
	=&\ 
	||\bs{S}^{-1}(\phi^0_1\bs{I}_n + \sum^{Q}_{q=1}\phi^0_{q+1}\bs{W}_{\kern -0.2em {q}}) - \bs{S}^{-1}(\bs{\hat{\rho}})(\hat{\phi}_1\bs{I}_n + \sum^{Q}_{q=1}\hat{\phi}_{q+1}\bs{W}_{\kern -0.2em q})||_2  \notag \\
	\leq&\ 
	||\bs{S}^{-1}(\phi^0_1\bs{I}_n + \sum^{Q}_{q=1}\phi^0_{q+1}\bs{W}_{\kern -0.2em {q}}) - \bs{S}^{-1}(\bs{\hat{\rho}})(\phi^0_1\bs{I}_n + \sum^{Q}_{q=1}\phi^0_{q+1}\bs{W}_{\kern -0.2em {q}})||_2 \notag \\
	&+\
	||\bs{S}^{-1}(\bs{\hat{\rho}})(\phi^0_1\bs{I}_n + \sum^{Q}_{q=1}\phi^0_{q+1}\bs{W}_{\kern -0.2em {q}}) - \bs{S}^{-1}(\bs{\hat{\rho}})(\hat{\phi}_1\bs{I}_n + \sum^{Q}_{q=1}\hat{\phi}_{q+1}\bs{W}_{\kern -0.2em q})||_2 \notag \\
	\leq&\ 
	||\bs{S}^{-1}||_2 ||(\bs{I}_n -  \bs{S}\bs{S}^{-1}(\bs{\hat{\rho}}))(\phi^0_1\bs{I}_n + \sum^{Q}_{q=1}\phi^0_{q+1}\bs{W}_{\kern -0.2em q})||_2 \notag \\
	&+\
	||\bs{S}^{-1}(\bs{\hat{\rho}})||_2 ||(\phi^0_1 - \hat{\phi}_1) \bs{I}_n + \sum^{Q}_{q=1}(\phi^0_{q+1} - \hat{\phi}_{q+1}) \bs{W}_{\kern -0.2em {q}})||_2 \notag \\
	\leq&\ 
	||\bs{S}^{-1}||_2 ||\bs{I}_n -  \bs{S}\bs{S}^{-1}(\bs{\hat{\rho}})||_2 ||\bs{S}||_2 ||\bs{A}||_2 \notag \\
	&+\
	||\bs{S}^{-1}(\bs{\hat{\rho}})||_2 (|\phi^0_1 - \hat{\phi}_1| \ + \sum^{Q}_{q=1} |\phi^0_{q+1} - \hat{\phi}_{q+1}| ||\bs{W}_{\kern -0.2em {q}}||_2) \notag \\
	\leq&\ 
	||\bs{S}^{-1}||_2 ||\bs{I}_n -  \bs{S}\bs{S}^{-1}(\bs{\hat{\rho}})||_2 ||\bs{S}||_2 ||\bs{A}||_2
	+
	||\bs{S}^{-1}(\bs{\hat{\rho}})||_2 ||\bs{\hat{\theta}} - \bs{\theta}^0||_2
	\notag \\
	&+\
 	||\bs{S}^{-1}(\bs{\hat{\rho}})||_2 ||\bs{\hat{\theta}} - \bs{\theta}^0||_2  \sqrt{Q} \max_{1 \leq q \leq Q} ||\bs{W}_{\kern -0.2em {q}}||_2 \notag \\
	=&\
	O_P(\sqrt{{Q}}||\bs{\hat{\theta}} - \bs{\theta}^0||_2),
\end{align}
where $||\bs{I}_n -  \bs{S}\bs{S}^{-1}(\bs{\hat{\rho}})||_2 = ||\bs{S}\bs{S}^{-1}(\bs{\hat{\rho}}) \bs{S}(\bs{\hat{\rho}})\bs{S}^{-1} -  \bs{S}\bs{S}^{-1}(\bs{\hat{\rho}})||_2 \leq ||\bs{S}\bs{S}^{-1}(\bs{\hat{\rho}}) ||_2 ||\bs{S}(\bs{\hat{\rho}})\bs{S}^{-1} -  \bs{I}_n||_2 = O_P(\sqrt{{Q}}||\bs{\hat{\theta}} - \bs{\theta}^0||_2)$ using Lemma \ref{L2.1} and Assumption \ref{AS:2.4}. The result then follows. 
\end{pff}

\begin{pff}[Proof of Proposition \ref{Prop6}] 
The proof largely follows the same structure as the proof of Theorem 3.5 in \cite{lu_shrinkage_2016}. Details can be found in Appendix D in the Supplementary Material. 
\end{pff}

\end{appendices}
 
%%%%%%%%%%    References   %%%%%%%%%%%
\bibliographystyle{elsart-harv}

\begin{thebibliography}{33}
	\expandafter\ifx\csname natexlab\endcsname\relax\def\natexlab#1{#1}\fi
	\expandafter\ifx\csname url\endcsname\relax
	\def\url#1{\texttt{#1}}\fi
	\expandafter\ifx\csname urlprefix\endcsname\relax\def\urlprefix{URL }\fi
	
	\bibitem[{Bai(2009)}]{bai_panel_2009}
	Bai, J., 2009. Panel data models with interactive fixed effects. Econometrica
	77~(4), 1229--1279.
	
	\bibitem[{Bai and Li(2021)}]{bai_dynamic_2017}
	Bai, J., Li, K., 2021. Dynamic spatial panel data models with common shocks.
	Journal of Econometrics 224~(1), 134--160.
	
	\bibitem[{Bai and Liao(2017)}]{bai_inferences_2017}
	Bai, J., Liao, Y., 2017. Inferences in panel data with interactive effects
	using large covariance matrices. Journal of Econometrics 200~(1), 59--78.
	
	\bibitem[{Bai and Ng(2002)}]{baiandng}
	Bai, J., Ng, S., 2002. Determining the number of factors in approximate factor
	models. Econometrica 70~(1), 191--221.
	
	\bibitem[{Blume et~al.(2015)Blume, Brock, Durlauf, and Jayaraman}]{blume}
	Blume, L.~E., Brock, W.~A., Durlauf, S.~N., Jayaraman, R., 2015. Linear social
	interactions models. Journal of Political Economy 123~(2), 444--496.
	
	\bibitem[{Bramoull\'e et~al.(2009)Bramoull\'e, Djebbari, and
		Fortin}]{bramoulle_identification_2009}
	Bramoull\'e, Y., Djebbari, H., Fortin, B., 2009. Identification of peer effects
	through social networks. Journal of Econometrics 150~(1), 41--55.
	
	\bibitem[{de~Paula et~al.(2020)de~Paula, Rasul, and
		Souza}]{de_paula_recovering_nodate}
	de~Paula, {\'A}., Rasul, I., Souza, P., 2020. Identifying network ties from
	panel data: Theory and an application to tax competition. Working paper,
	{CeMMAP}.
	
	\bibitem[{Durlauf et~al.(2009)Durlauf, Johnson, and Temple}]{newgrow}
	Durlauf, S.~N., Johnson, P.~A., Temple, J.~R., 2009. The methods of growth
	econometrics. In: Mills, T., Patterson, K. (Eds.), Palgrave Handbook of
	Econometrics. Vol.~2. Elsevier, Ch.~8, pp. 1119--1179.
	
	\bibitem[{Fan and Peng(2004)}]{peng_nonconcave_2004}
	Fan, J., Peng, H., 2004. Nonconcave penalized likelihood with a diverging
	number of parameters. The Annals of Statistics 32~(3), 928--961.
	
	\bibitem[{Gupta and Robinson(2015)}]{gupta_inference_2015}
	Gupta, A., Robinson, P.~M., 2015. Inference on higher-order spatial
	autoregressive models with increasingly many parameters. Journal of
	Econometrics 186~(1), 19--31.
	
	\bibitem[{Gupta and Robinson(2018)}]{gupta_pseudo_2018}
	Gupta, A., Robinson, P.~M., 2018. Pseudo maximum likelihood estimation of
	spatial autoregressive models with increasing dimension. Journal of
	Econometrics 202~(1), 92--107.
	
	\bibitem[{Horn and Johnson(2012)}]{Horn}
	Horn, R.~A., Johnson, C.~R., 2012. Matrix Analysis, 2nd Edition. Cambridge
	University Press, New York, USA.
	
	\bibitem[{Hsiao(2018)}]{hsiao_panel_2018}
	Hsiao, C., 2018. Panel models with interactive effects. Journal of Econometrics
	206~(2), 645--673.
	
	\bibitem[{Kuersteiner and Prucha(2020)}]{prucha}
	Kuersteiner, G.~M., Prucha, I.~R., 2020. Dynamic spatial panel models:
	Networks, common shocks, and sequential exogeneity. Econometrica 88~(5),
	2109--2146.
	
	\bibitem[{Lam and Souza(2019)}]{Lam_estimation_2019}
	Lam, C., Souza, P.~C., 2019. Estimation and selection of spatial weight matrix
	in a spatial lag model. Journal of Business \& Economic Statistics 38~(3),
	693--710.
	
	\bibitem[{Latala(2005)}]{latala_estimates_nodate}
	Latala, R., 2005. Some estimates of norms of random matrices. Proceedings of
	the American Mathematical Society 133~(5), 1273--1282.
	
	\bibitem[{Lee(2007)}]{lee_identification_2007}
	Lee, L.-F., 2007. {Identification and estimation of econometric models with
		group interactions, contextual factors and fixed effects}. Journal of
	Econometrics 140~(2), 333--374.
	
	\bibitem[{Lee et~al.(2010)Lee, Liu, and Lin}]{lee_specification_2010}
	Lee, L.-F., Liu, X., Lin, X., 2010. Specification and estimation of social
	interaction models with network structures. The Econometrics Journal 13~(2),
	145--176.
	
	\bibitem[{Lee and Yu(2010)}]{lee_estimation_nodate}
	Lee, L.-F., Yu, J., 2010. Estimation of spatial autoregressive panel data
	models with fixed effects. Journal of Econometrics~(1), 165--185.
	
	\bibitem[{Lee and Yu(2014)}]{SpatialPanelDataModels}
	Lee, L.-F., Yu, J., 2014. Spatial panel data models. In: Baltagi, B.~H. (Ed.),
	The {O}xford Handbook of Panel Data. {Oxford} {University} {Press}, Ch.~12,
	pp. 363--401.
	
	\bibitem[{Leeb and P\"{o}tscher(2005)}]{leeb}
	Leeb, H., P\"{o}tscher, B.~M., 2005. Model selection and inference: Facts and
	fiction. Econometric Theory 21~(1), 21--59.
	
	\bibitem[{Lewbel et~al.(2021)Lewbel, Qu, and Tang}]{lewbel_2019}
	Lewbel, A., Qu, X., Tang, X., 2021. Social networks with unobserved links.
	Working paper.
	
	\bibitem[{Liu(2017)}]{liu_doctor_2017}
	Liu, T., 2017. Model selection and adaptive lasso estimation of spatial models.
	{PhD} {Thesis}, {The} {Ohio} {State} {University}.
	
	\bibitem[{Lu and Su(2016)}]{lu_shrinkage_2016}
	Lu, X., Su, L., 2016. Shrinkage estimation of dynamic panel data models with
	interactive fixed effects. Journal of Econometrics 190~(1), 148--175.
	
	\bibitem[{Moon and Weidner(2015)}]{moon_linear_2015}
	Moon, H.~R., Weidner, M., 2015. Linear regression for panel with unknown number
	of factors as interactive fixed effects. Econometrica 83~(4), 1543--1579.
	
	\bibitem[{Moon and Weidner(2017)}]{moon_dynamic_nodate}
	Moon, H.~R., Weidner, M., 2017. Dynamic linear panel regression models with
	interactive fixed effects. Econometric Theory 33~(1), 158--195.
	
	\bibitem[{Newey and McFadden(1994)}]{Newey}
	Newey, W.~K., McFadden, D., 1994. Large sample estimation and hypothesis
	testing. In: Engle, R.~F., McFadden, D.~L. (Eds.), Handbook of Econometrics.
	Vol.~4. Elsevier, Ch.~36, pp. 2111--2245.
	
	\bibitem[{Shi and Lee(2017)}]{shi_spatial_2017}
	Shi, W., Lee, L.-F., 2017. Spatial dynamic panel data models with interactive
	fixed effects. Journal of Econometrics 197~(2), 323--347.
	
	\bibitem[{Shi and Lee(2018)}]{endo}
	Shi, W., Lee, L.-F., 2018. A spatial panel data model with time varying
	endogenous weights matrices and common factors. Regional Science and Urban
	Economics 72, 6--34.
	
	\bibitem[{Wang(2018)}]{yike}
	Wang, Y., 2018. Panel data with high-dimensional factors: Inference on
	treatment effects with an application to sampled networks. Working paper.
	
	\bibitem[{Yu et~al.(2008)Yu, de~Jong, and Lee}]{yu_quasi-maximum_2008}
	Yu, J., de~Jong, R., Lee, L.-F., 2008. Quasi-maximum likelihood estimators for
	spatial dynamic panel data with fixed effects when both $n$ and {$T$} are
	large. Journal of Econometrics 146~(1), 118--134.
	
	\bibitem[{Zhang and Yu(2018)}]{zhang_spatial_2018}
	Zhang, X., Yu, J., 2018. Spatial weights matrix selection and model averaging
	for spatial autoregressive models. Journal of Econometrics 203~(1), 1--18.
	
	\bibitem[{Zou(2006)}]{zou_adaptive_2006}
	Zou, H., 2006. The adaptive lasso and its oracle properties. Journal of the
	American Statistical Association 101~(476), 1418--1429.
\end{thebibliography}

\end{document}